%% file: main.tex
\documentclass[%
reprint,
amsmath,amssymb,
aps,
pra,
]{revtex4-2}

\usepackage{xcolor}
\usepackage{graphicx}% Include figure files
\usepackage{dcolumn}% Align table columns on decimal point
\usepackage{bm}% bold math
\usepackage{orcidlink}
\usepackage{mathtools}
\usepackage{physics}

\usepackage{tikz}
\usepackage{quantikz}
\usetikzlibrary{shapes.geometric}
\usetikzlibrary{arrows.meta}

\usepackage{dsfont} % For a nice identity \mathds{1}

\usepackage{soul}

\usepackage{hyperref}
\hypersetup{
     colorlinks   = true,
     linkcolor    = red,
     citecolor    = blue,
     urlcolor     = blue,
     }
\usepackage[capitalize]{cleveref}

\newcommand{\ii}{\ensuremath{\mathrm{i}}} % Imaginary unit
\newcommand{\id}{\ensuremath{\mathbb{I}}} % Identity
\newcommand{\Proj}[1]{\big[\!\big[#1\big]\!\big]}
\newcommand{\kket}[1]{| #1 \rangle\!\rangle}
\newcommand{\bbra}[1]{\langle\!\langle #1 |}
\begin{document}

\preprint{APS/123-QED}

\title{Benchmarking indirect quantum control schemes via higher-order quantum operations}% Force line breaks with \\

\author{\v{S}imon Vedl$^{1}$ \orcidlink{0009-0006-6575-9941}}
\thanks{\href{mailto:simon.vedl@hdr.mq.edu.au}{simon.vedl@hdr.mq.edu.au};
\href{mailto:varun.srivastava@hdr.mq.edu.au}{varun.srivastava@hdr.mq.edu.au}.\\
SV and VS contributed equally.}

\author{Varun Srivastava$^{1}$ \orcidlink{0000-0002-3907-5304}}
\thanks{\href{mailto:simon.vedl@hdr.mq.edu.au}{simon.vedl@hdr.mq.edu.au};
\href{mailto:varun.srivastava@hdr.mq.edu.au}{varun.srivastava@hdr.mq.edu.au}.\\
SV and VS contributed equally.}

\author{Riddhi Ghosh$^{1}$ \orcidlink{0000-0003-0856-5412}}

\author{Alexei Gilchrist$^{1}$ \orcidlink{0000-0003-0075-5174}}

\affiliation{$^{1}$School of Physical and Mathematical Sciences, Macquarie University, Sydney NSW, Australia}%

\date{\today}% It is always \today, today,

\begin{abstract}
Indirect quantum control aims to manipulate a target system through interventions on an auxiliary controller with which it interacts. We formulate finite-step indirect-control protocols using the language of higher-order quantum operations, representing admissible controller manipulations as deterministic superinstruments. This representation separates the uncontrolled but fixed dynamics from the controllable operations and allows the optimal control problem to be written as a semidefinite program for a chosen figure of merit. The resulting optimum gives an operational benchmark by providing the best performance achievable by a finite sequence of control operations, including protocols with classical or quantum feed-forward. More restricted and experimentally motivated control classes, such as independent unitary controls or memoryless quantum channels, can then be compared against this benchmark. We illustrate the framework with a two-step qubit-purification task, in which an initially mixed target qubit is steered toward a pure state through a fixed interaction with a controller qubit. The example shows regimes where simple unitary strategies saturate the higher-order benchmark, as well as regimes where they are provably suboptimal. This approach provides a systematic way to study the value of control resources in indirect quantum-control schemes. 
\end{abstract}

\maketitle

%\tableofcontents

\section{Introduction}\label{sec:Introduction}

Quantum control concerns the manipulation of dynamics of quantum systems through suitably designed control operations. An important problem in quantum control is the steering of the quantum systems towards desired target states. Since the advent of quantum mechanics, the control of quantum systems has played a central role in both foundational studies and practical applications \cite{Warren1993,Chu2002-ua,blaquiere2014information,wiseman2009quantum,1983differential}. More recently, it has become an essential ingredient in quantum computation \cite{Burgarth2009,Burgarth2010,Morton2006,Hodges2008}, metrology \cite{Degen2017}, and error mitigation \cite{Taminiau2014,matsos2025}. For isolated systems under coherent control, the question of controllability is typically determined by the dynamics generated by the available set of control Hamiltonians. In particular, under standard Lie-algebraic conditions, repeated application of a finite set of controls can generate full controllability over the system dynamics \cite{1983differential,Peirce1988,Huang1983}.

In many experimental settings, however, the system of interest is not the most convenient degree of freedom to address directly. Direct driving may be impractical, slow, or experimentally suboptimal. A prominent example arises in platforms involving hybrid quantum systems consisting of electron-nuclear spins, such as NV centres in diamond, where it is often preferable to control the electron spin rather than the nuclear spin. The electron spin couples to nearby nuclear spins through hyperfine interactions, allowing faster gate implementations than would typically be possible through direct nuclear-spin control \cite{Hodges2008,Zhang_2019,Zhang_2011,Taminiau2014,Liu2013}. Similarly, in systems of trapped ions, the spin can be coupled to the mechanical motion degrees of freedom which allows controlling the motion state of the ion by acting on its spin \cite{matsos2025}. This motivates the paradigm of indirect control: given a target system $S$ coupled to a quantum controller $C$, to what extent can a restricted set of operations on $C$ be leveraged to implement and optimise control over $S$? This question has been studied across several experimental platforms, including spin chains \cite{Burgarth2009,Burgarth2010} and coupled electron-nuclear spin registers \cite{Hodges2008,Morton2006,Cappellaro2009}. On the theoretical side, a range of studies have established conditions for indirect controllability and introduced broader mathematical frameworks for analysing actuator-mediated control \cite{Lloyd2001,Lloyd2004,Burgarth2007,DAlessandro2012,Layden2016}. In this paper, we use the terms controller and actuator interchangeably.

Here we present a complementary perspective on indirect quantum control by formulating it within the framework of higher-order quantum operations (HOQOs). In an indirect-control protocol, the experimentally accessible objects are not simply state transformations, but interventions inserted at different times into a larger fixed dynamical process. HOQOs provide a natural language for such settings because they describe transformations acting on quantum operations themselves \cite{taranto2025,Chiribella2009}. HOQOs have been used extensively to investigate causal structures in quantum mechanics \cite{Oreshkov2012,Costa_2016}, to provide operational characterisations of quantum non-Markovianity \cite{Pollock2018,Milz2021,Giarmatzi_2021}, and are increasingly finding practical applications in benchmarking, error mitigation, and quantum control \cite{White2020,Giarmatzi2025,Figueroa-Romero2021,srivastava2025,tanggara2024,White2025,Zambon2025,roy2026}. They have also been used to provide operational methods for characterising the performance of a complex quantum networks \cite{Chiribella_2016}. In our approach, the admissible control protocols on the controller are modelled as HOQOs, specifically as superinstruments \cite{taranto2025}. Superinstruments have appeared in the literature under several names, including quantum strategies \cite{Gutoski2007} and testers \cite{Ziman2008}. 

In control theory, questions of controllability are often posed as existence problems, i.e., given fixed dynamics and a desired target, does there exist a control protocol that achieves the task? Such results identify when control is possible in principle. However, when we restrict to control sequences of fixed length, the representations in terms of superinstruments allows us to recast the existence problem as a semidefinite program (SDP). The solution of the SDP provides an optimal superinstrument and even when the superinstrument does not admit a simple physical implementation, its performance still furnishes a fundamental upper bound on what can be achieved using finite time quantum interventions on the controller. A key advantage of the SDP formulation is therefore, that it provides a rigorous benchmark for indirect quantum control. Consequently, more restricted control strategies such as protocols based solely on unitary operations or limited ancilla resources can be quantitatively compared against this benchmark. The HOQO formulation is also highly modular and additional physical constraints can be incorporated by modifying the underlying process tensor while leaving the optimisation framework unchanged. For example, the same approach naturally extends to settings in which the target system is subject to decoherence or environmental interactions \cite{Owari2015}, thereby providing a unified framework for studying optimal indirect control in both closed and open quantum systems.

Specifically, we consider the largest class of deterministic protocols compatible with our setting, i.e., a fixed interaction between the target system $S$ and controller $C$, together with a finite number of discrete intervention times at which quantum operations are applied to the controller. Importantly, the admissible protocols encompass a broad range of indirect-control primitives previously studied in the literature. These include trace-and-replace operations applied at intermediate times \cite{Layden2016}, SWAP-based control mechanisms enabled by ancillary systems coupled to the controller \cite{Burgarth2007}, and more general temporally correlated interventions capable of generating memory effects and implementing classical or quantum feed-forward across multiple time steps.

In this work we begin by briefly introducing the framework of higher-order quantum operations (HOQOs) and using it to describe the problem of indirect control. We show that using the properties of the link product the dynamics can be conveniently partitioned into state preparation, fixed dynamics, control operations, and final readout. We then show that a sequence of control operations can be represented by a deterministic superinstrument, which is a semidefinite operator subject to linear constraints that ensure the causal ordering of interventions. In \cref{sec:IndirectControl}, we show how the problem of finding an optimal control sequence can be framed as a semidefinite program (SDP). Finally, we showcase the formalism by applying it to the task of two-step qubit purification, where we demonstrate the benchmarking power of this framework.

\section{Indirect control using HOQOs: Setup}
\label{sec:HOQO}

Indirect control schemes involve applying a sequence of quantum operations to an actuator in order to steer a target system that may not be directly addressable. In the most general setting, the target system and the controller may also interact with additional degrees of freedom, so that the resulting dynamics is naturally described as a multi-time process. HOQOs provide a unified language for such scenarios. While ordinary quantum operations act on quantum states—for example, measurements map states to probability distributions and channels map input states to output states—higher-order operations transform quantum operations into other quantum operations~\cite{taranto2025,Chiribella2009}. This hierarchy of ``operations acting on operations'' is particularly well suited to model indirect control schemes as we outline next.

We consider a target system $S$ that is not directly accessible. Instead, $S$ interacts with a controller system $C$ through a fixed joint dynamics. This interaction is assumed to be well characterised, as is often the case in platforms such as NV centres in diamond \cite{DOHERTY20131}. The controller $C$ can be accessed at discrete time steps, and it can be potentially coupled to an additional ancilla system $A$. The aim is to exploit this restricted access, together with the fixed $S$-$C$ interaction, to steer the target system $S$ towards a desired final state. Within the HOQO formalism, the admissible multi-time control protocols are represented by deterministic superinstruments, which encode the most general deterministic sequence of interventions available on the controller degrees of freedom.

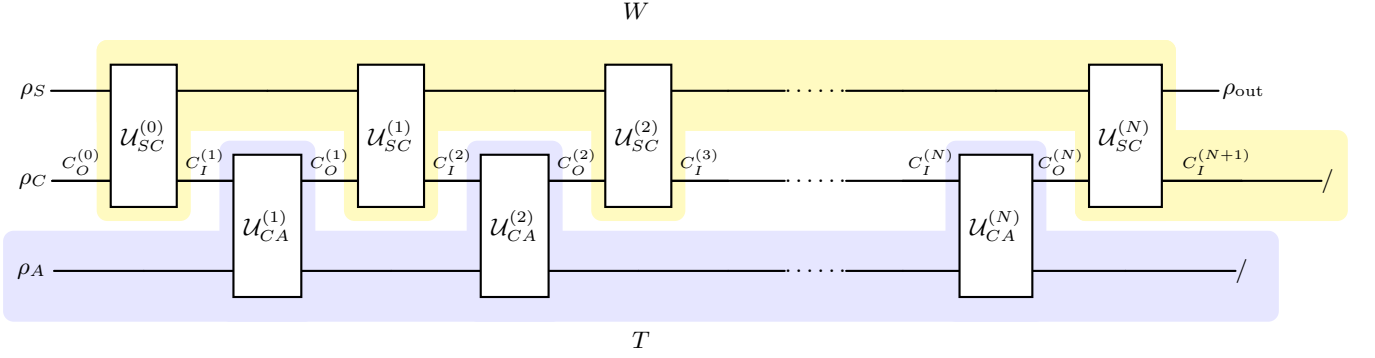
\begin{figure*}[t]
  \centering
  \input{Figures/Tikz/process-tester-comb.tex}
  \caption{Illustration of how the dynamics between the system, controller, and ancilla can be separated into state preparation, process tensor $W$ corresponding to fixed dynamics, and deterministic superinstrument $T$ representing the control operations altogether combining to produce the final output state $\rho_{\mathrm{out}}$.}
  \label{fig:ProcessTesterCombs}
\end{figure*}

In the following, we focus on HOQOs embedded in a fixed causal structure. Such objects can be built from elementary quantum operations, such as completely positive maps, by specifying how the output spaces of some operations are connected to the input spaces of others. A convenient way to describe this composition is to use the Choi--Jamio\l{}kowski (CJ) representation of quantum maps~\cite{JAMIOLKOWSKI1972,CHOI1975}, together with the link product for composing the corresponding CJ operators~\cite{Chiribella_2016,taranto2025}. For a completely positive, linear map $\mathcal{M}:\mathcal{L}(A)\to\mathcal{L}(B)$, the CJ operator is the positive semidefinite operator $\Proj{\mathcal{M}}$ defined by:
\begin{equation}
\Proj{\mathcal{M}} := (\mathcal{I}\otimes\mathcal{M})(\kket{\mathbb{I}}\bbra{\mathbb{I}})
\in \mathcal{L}(A\otimes B),
\end{equation}
where $\kket{\mathbb{I}}=\sum_j \ket{j}\otimes\ket{j}$ is the unnormalised maximally entangled state in a fixed basis. We consider a process with $N$ control operations applied to the controller $C$. At time $t=0$, the systems are initialised in states $\rho_S\in \mathcal{L}(S^{(0)})$, $\rho_C\in \mathcal{L}(C_O^{(0)})$, and $\rho_A\in \mathcal{L}(A^{(0)})$. Here, the superscript $(t)$ labels the time step. At each time step $t\in\{0,\dots,N\}$, the target system $S$ interacts with the controller $C$ through a fixed joint unitary $\mathcal{U}_{SC}^{(t)}$. In the CJ representation, its Choi operator lies in the Hilbert space $\mathcal{L}\!\bigl(S^{(t)}\otimes C_O^{(t)}\otimes S^{(t+1)}\otimes C_I^{(t+1)} \bigr)$ where the subscripts $I$ and $O$ denote the input and output spaces of the controller associated with each control step (see \cref{fig:ProcessTesterCombs}). Between successive $S$--$C$ interactions, we apply joint unitaries $\mathcal{U}_{CA}^{(t)}$ on the controller and ancilla systems for $t\in\{1,\dots,N\}$, representing the finite time-step control operation, and their CJ operators lie in Hilbert space $\mathcal{L}\!\bigl(C_I^{(t)}\otimes A^{(t-1)}\otimes C_O^{(t)}\otimes A^{(t)}\bigr)$. The two classes of interactions of the controller are therefore interleaved according to the ordering
\[
C_O^{(t)}
\;\xrightarrow{\;\mathcal{U}_{SC}^{(t)}\;}\;
C_I^{(t+1)}
\;\xrightarrow{\;\mathcal{U}_{CA}^{(t+1)}\;}\;
C_O^{(t+1)}.
\]
We assume full control over the $C$--$A$ operations, while the $S$--$C$ interactions are fixed and not under our control. The objective is to determine whether there exists a sequence of admissible $C$--$A$ operations that steers the target system $S$ to a desired final state $\rho_{t}\in\mathcal{L}(S^{(N+1)})$.

In order to compose maps in the CJ representation, we use the \emph{link product}. Let $M\in\mathcal{L}(P)$ and $N\in\mathcal{L}(Q)$, where the Hilbert spaces $P$ and $Q$ may share a common tensor factor $P\cap Q$. Their link product is defined as
\begin{equation}
M\star N
=
\Tr_{P\cap Q}
\!\left[
\bigl(\id^{Q\setminus P}\otimes M^{T_{P\cap Q}}\bigr)
\bigl(N\otimes \id^{P\setminus Q}\bigr)
\right],
\label{eq:linkprod}
\end{equation}
where $T_{P\cap Q}$ denotes the partial transpose on the shared space, and $\Tr_{P\cap Q}$ represents the partial trace over the shared space. The link product preserves Hermiticity and positivity, and is associative provided the Hilbert-space labels are tracked consistently \cite{taranto2025}. 

The final output state is given by the link product of the form (See \cref{app:Linkprod} for details)

\begin{equation}
\rho_{\mathrm{out}}
=
\rho_S \otimes \rho_C \star W \star T.
\label{eq:final_output}
\end{equation}
Here,
\begin{equation}
W=
\Proj{\mathcal{U}_{SC}^{(0)}}
\star
\Proj{\mathcal{U}_{SC}^{(1)}}
\star\cdots\star
\Proj{\mathcal{U}_{SC}^{(N)}}
\star
\id^{C_I^{(N+1)}}
\label{eq:process_tensor}
\end{equation}

is the process tensor, which encodes the multi-time dynamics of the uncontrolled $S$--$C$ interactions \cite{Pollock2018,Giarmatzi_2021}. Note that the last $\id^{C_I^{(N+1)}}$ operation implies that we are discarding the final state of the controller. Process tensors map a sequence of control operations to a final output state. By construction it is also a positive semi-definite operator.

The control protocol acting on the controller--ancilla degrees of freedom is represented by a deterministic superinstrument

\begin{equation}
T
=
\rho_A \star \Proj{\mathcal{U}_{CA}^{(1)}}
\star
\Proj{\mathcal{U}_{CA}^{(2)}}
\star\cdots\star
\Proj{\mathcal{U}_{CA}^{(N)}}
\star
\id^{A^{(N)}}.
\label{eq:superinstrument_def}
\end{equation}

the operator $T$ lies in the Hilbert space $\mathcal{L}\!\left(\bigotimes_{t=1}^{N}C_I^{(t)}\otimes C_O^{(t)}\right)$. Thus, $T$ has one input leg $C_I^{(t)}$ and one output leg $C_O^{(t)}$ for each time step $t$. More generally, a superinstrument is specified by a collection of positive semidefinite operators $\{T_a\}_a$, where each element corresponds to a particular outcome $a$. When linked with the process tensor using eq.~\eqref{eq:final_output}, $T_a$ produces an output state $\rho_a$ with probability $p_a$, thereby describing a non-deterministic control process. The corresponding deterministic superinstrument is obtained by coarse-graining over outcomes, so that $\sum_a T_a=T$.

A deterministic superinstrument is not an arbitrary positive operator. Its CJ operator must satisfy a hierarchy of linear constraints expressing causality, or equivalently, the absence of signalling from future time steps to past ones \cite{taranto2025}. Defining

\begin{equation}
\mathcal{P}_{X}(A)
:=
\Tr_{X}(A)\otimes \frac{\id^{X}}{d^{X}},
\label{eq:replace_map}
\end{equation}
where $d^X$ is the dimension of the Hilbert space $X$, the final causality constraint can be written as
\begin{equation}
\mathcal{P}_{C_O^{(N)}}({T})
=
\mathcal{P}_{C_I^{(N)}}\mathcal{P}_{C_O^{(N)}}({T}).
\label{eq:last_constraint}
\end{equation}

We can re-write \cref{eq:last_constraint} by introducing the linear operator
\[
\mathcal{L}_{N}(T)
=
\mathcal{P}_{C_I^{(N)}}\mathcal{P}_{C_O^{(N)}}({T})
-
\mathcal{P}_{C_O^{(N)}}({T})=0,
\]
More generally, for $n=1,\dots,N$, the full hierarchy of causality constraints can be written as
\begin{equation}
\mathcal{L}_{n}(T)=0
\label{eq:general_constraint}
\end{equation}
where $\mathcal{L}_{n}(T)=\mathcal{P}_{C_I^{(n)}}\mathcal{P}_{C_O^{(n)}}\ldots \mathcal{P}_{C_I^{(N)}}\mathcal{P}_{C_O^{(N)}}({T})-\mathcal{P}_{C_O^{(n)}} \ldots \mathcal{P}_{C_I^{(N)}}\mathcal{P}_{C_O^{(N)}}({T})$. Together with the positivity condition $T\geq 0$ and the normalisation constraint $\Tr(T)=D_{I}$, these constraints characterise valid deterministic superinstruments. Here $D_I=\prod_{i=1}^{N}d^{C_I^{(i)}}$ is the product of the dimensions of all input Hilbert spaces of the controller.

\section{Benchmarking indirect control schemes}
\label{sec:IndirectControl}
The SDP formulation gives two advantages. First, it provides a bound on the optimal achievable performance for a given task. Second, whenever the optimum is attained, it returns an explicit control protocol, represented by the deterministic superinstrument $T$, that realises this bound. To formulate the problem as an SDP we note that \cref{eq:final_output} gives the final state of the system in terms of a fixed process tensor $W$ and the deterministic superinstrument  $T$. The initial states $\rho_S$ and $\rho_C$, together with the process tensor $W$, are fixed data of the problem. The optimisation variable is the control object $T$. Furthermore, for a broad class of information-processing tasks, the figure of merit is a real linear function of the output state and can therefore be written as $\Tr[\Omega\,\rho_{\mathrm{out}}]$ for some Hermitian operator $\Omega$ acting on the final Hilbert space of the target system. Since the link product is linear in each of its arguments, the objective $\Tr[\Omega\,\rho_{\mathrm{out}}]$ is linear in $T$. Moreover, the admissibility conditions on $T$ derived in the previous section consist of positivity, normalisation, and linear causality constraints. Therefore, the problem of optimal indirect control can be expressed as the following SDP:
\begin{equation}
  \label{eq:sdp_indirect_control}
  \begin{aligned}
  \max \quad
  & \Tr\!\Big[
  \Omega \big(
  \rho_S\otimes\rho_C
  \star W \star T
  \big)
  \Big] \\
  \text{s.t.}\quad
  & T\geq 0, \\
  & \mathcal{L}_{n}(T)=0\quad\forall n=1,\dots,N , \\
  & \Tr(T)=D_I
  \end{aligned}
\end{equation}

This optimisation runs over all admissible multi-time control sequences. Solving the SDP provides two results: first, it determines the optimal achievable performance and hence certifies whether a specified target performance is attainable, and second, it provides the optimal control sequence represented by the deterministic superinstrument $T$. More importantly, its optimal value also provides a benchmark for any class of deterministic control sequences (see \cref{app:DualSDP}). In particular, even when the optimal deterministic superinstrument is not easily synthesised into a simple circuit or pulse-level description, the SDP optimum gives an upper bound against which physically restricted ans\"atze, such as unitary-only control operations, can be compared, as we show in \cref{sec:Results}.

A useful feature of the HOQO description is that it naturally separates the fixed dynamics from the admissible interventions. Additional physical structure can therefore be incorporated into the fixed process without changing the overall optimisation logic. For example, one may enlarge $W$ to account for decoherence or dissipation acting on the target system $S$ between intervention times, or to include more general environmental memory effects \cite{Owari2015}. The same SDP framework then applies, with the modified physical assumptions absorbed into the process tensor, while the admissibility constraints on $T$ remain unchanged. To make this explicit, consider an environment $E$ that interacts continuously with the target system $S$. The environmental influence can be incorporated by replacing the fixed unitary interactions with a larger family of joint unitaries $\mathcal{U}^{(t)}_{ESC}$ acting on the environment $E$, the system $S$, and the controller degrees of freedom $C$. The corresponding process tensor is then given by $W=\rho_{SE}\star\Proj{\mathcal{U}_{ESC}^{(0)}}\star\Proj{\mathcal{U}_{ESC}^{(1)}}\star\cdots\star\Proj{\mathcal{U}_{ESC}^{(N)}}\star\id^{C_I^{(N+1)}}\star\id^{E^{(N+1)}}$. Thus, decoherence, dissipation, or memory effects associated with $E$ are incorporated directly into the process tensor $W$.

One may also enlarge the optimisation problem by allowing the initial resources to vary. In \cref{eq:final_output}, the initial states of the relevant systems are treated as fixed data. However, if one is additionally free to choose the initial states of $S$, $C$, and any auxiliary ancilla $A$, then these can be promoted to optimisation variables. For fixed $W$ and fixed $T$, the final output state remains linear in the chosen initial state, and therefore any objective of the form $\Tr[\Omega \rho_{\mathrm{out}}]$ is also linear in that variable. This opens the possibility of optimising not only over control protocols, but also over the best preparable initial resources compatible with the experimental setting. More generally, one can consider alternating optimisations over initial states and superinstruments, for example by using a see-saw SDP.

\section{Example: 2-step qubit polarisation}\label{sec:Results}

In this section, we use a polarisation task to illustrate the benchmarking role of the framework developed thus far for a two time-step process. Our aim is to compare the optimal performance achievable by a general deterministic superinstrument to the one obtained from a more restricted and experimentally motivated class of control protocols consisting of independent unitary operations applied to the controller at each intervention time.

\begin{figure}[h]
  \centering
  \input{Figures/Tikz/purification-diagram.tex}
  \caption{Circuit diagram representing the indirect control in the two-step purification process. We split the dynamics into two parts, the fixed elements --- the process tensor $W$ and state preparation, and the deterministic superinstrument $T$ capturing the operations on the controller.}
  \label{fig:PurificationCircuit}
\end{figure}
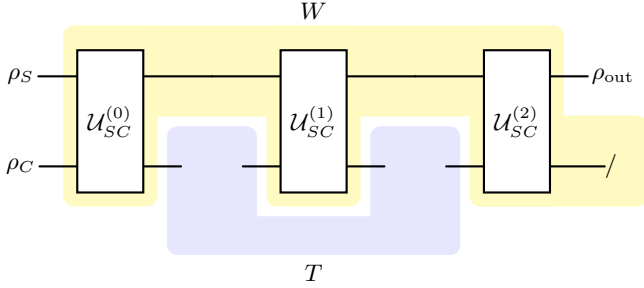

We consider a target qubit that is initially maximally mixed and interacts with an accessible controller qubit through a fixed interaction Hamiltonian. The controller is acted upon at two intermediate times, while the target is never directly controlled. Specifically, the two qubits first interact for a duration $t_0$, after which the first control intervention is applied to the controller. They then interact for a duration $t_1$, followed by a second control intervention, and finally evolve jointly for a duration $t_2$. At the end of the protocol, the controller is discarded and the target qubit is steered to its final state, as illustrated in \cref{fig:PurificationCircuit}. Our task is to polarise the target to $\ketbra{0}{0}$ state.

We partition the dynamics as indicated by \cref{eq:final_output} into initial state $\rho_S\otimes\rho_C$, the process tensor $W$ and the deterministic superinstrument $T$. The process tensor $W$ captures the dynamics between the system and the control qubit and can be expressed in terms of the link product of the Choi matrices of S-C unitaries

\begin{equation}
  W
  =\Proj{\mathcal{U}_{SC}^{(0)}} \star \Proj{\mathcal{U}_{SC}^{(1)}} \star \Proj{\mathcal{U}_{SC}^{(2)}} \star \id^{C_{I}^{(3)}},
  \label{eq:ExampleProcessMatrix}
\end{equation}
The system--controller unitaries ${\mathcal{U}_{SC}^{(i)}=\exp(\ii t_iH)}$ are generated by a fixed two-qubit Hamiltonian $H$. We consider two qubits with nearly matched effective transition frequencies, coupled through an anisotropic spin--spin interaction
\begin{equation}
  \begin{aligned}
    H ={}& \frac{\omega}{2}\,\sigma^{(S)}_z
    + \frac{(\omega+\Delta\omega)}{2}\,\sigma^{(C)}_z \\
    &+ J_{\parallel}\,\sigma^{(S)}_z\sigma^{(C)}_z \\
    &+ J_{\perp}\!\left(
      \sigma_{+}^{(S)}\sigma_{-}^{(C)}
      + \sigma_{-}^{(S)}\sigma_{+}^{(C)}
    \right).
  \end{aligned}
  \label{eq:Hamiltonian}
\end{equation}
Here, $\omega$ sets the common splitting of the two qubits, while $\Delta\omega$ denotes the residual detuning between the target
and controller transition frequencies. Parameter $J_{\parallel}$ is the longitudinal coupling and does not directly exchange populations but instead generates state-dependent phases and correlations between the two qubits. The exchange of population between the two qubits is generated by the transverse flip-flop interaction, with coupling strength $J_{\perp}$, where $\sigma_{\pm}=(\sigma_x\pm \ii\sigma_y)/2$. We emphasise that \cref{eq:Hamiltonian} is not intended to provide a microscopic model of any particular experimental platform. Rather, we use \cref{eq:Hamiltonian} as a simple and physically motivated testbed for illustrating the benchmarking framework and Hamiltonians of this form arise in a variety of physical settings. For example, driven electron--nuclear spin systems associated with nitrogen-vacancy centres in diamond can be engineered to realise effective flip-flop interactions of this form with a finite residual detuning
\cite{London_2013,Schwartz_2018,Whaites_2023}. Closely related exchange Hamiltonians also arise for capacitively coupled superconducting qubits, where the excitation-conserving interactions are exploited to apply $iSWAP$ gates \cite{Krantz_2019}.

The goal is to steer the system qubit towards the target state $\rho_t=\ketbra{0}{0}$. We evaluate the performance using the overlap of the final state with the target state $\bra{0}\rho_{\mathrm{out}}\ket{0}$ which can be cast as a linear function of the deterministic superinstrument $T$
\begin{equation}
  f(T) = \bra{0}(\rho_S\otimes\rho_C\star W \star T)\ket{0}.
\end{equation}
We assume that the target system is initialised in the maximally mixed state $\rho_S=\frac12\id$ which can denote a qubit at thermal equilibrium at room temperature. When we initialise the controller in the state $\rho_C=\ketbra{0}{0}$, the free evolution, i.e. no interventions, reads (See \cref{app:free_evolution})
\begin{equation}
  \label{eq:free_evol}
  f(\mathcal{I}) = \frac{1}{2} + \frac{1}{2} \left[\frac{1}{1+\delta^{2}} \operatorname{sin}^{2}\left(J_{\perp} \tau\sqrt{1+\delta^{2}}\right)\right]
\end{equation}
where $\tau=t_0+t_1+t_2$ and $\delta=\Delta \omega/2J_{\perp}$.

The relative magnitude of the transverse coupling and the residual detuning therefore determines the efficiency of the polarisation task. In particular, the free dynamics naturally separates into three regimes. When $|\Delta\omega|\ll 2J_{\perp}$, the two qubits are close to resonance and the transverse interaction can efficiently exchange population between the system and the controller. When $|\Delta\omega|\sim 2J_{\perp}$, the detuning and exchange interaction compete on comparable scales and the amount of population transfer becomes strongly dependent on both the interaction time and the precise value of the detuning. Finally, in the regime $|\Delta\omega|\gg 2J_{\perp}$, the two qubits are far off resonance and population exchange is suppressed. These regimes are naturally captured by the dimensionless ratio $\delta=\Delta\omega/(2J_{\perp})$. In a physical implementation, the longitudinal and transverse couplings generally originate from different components of the underlying anisotropic system--controller interaction. Their relative magnitude depends on the microscopic details of the platform, for example on the orientation and position of a nuclear spin relative to an electronic spin in anisotropic hyperfine interactions. Here we choose to set $J_{\parallel}=J_{\perp}\equiv J$, which represents a regime in which the longitudinal and transverse components of the interaction are of comparable strength. We note that the absence of $J_{\parallel}$ from \cref{eq:free_evol} is specific to the chosen initial states and the free evolution setting. In the presence of intermediate control operations, the longitudinal interaction can in general affect the dynamics.

The remaining dynamical scale is set by the total interaction time $\tau$. Since an overall rescaling of $J$, $\Delta\omega$, and $\tau$ leaves the dynamics invariant when the products $J\tau$ and $\Delta\omega\tau$ are fixed, we work in units of the coupling strength and set $J=1$. The two controller interventions are placed symmetrically within this interval, such that $t_0=t_1=t_2=\tau/3$. We then vary the detuning $\Delta\omega$, or equivalently the dimensionless quantity $\delta=\Delta\omega/(2J)$, thereby moving continuously from the near resonant regime to the far detuned regime.

\begin{figure}[h]
  \centering
  \includegraphics[width=\columnwidth]{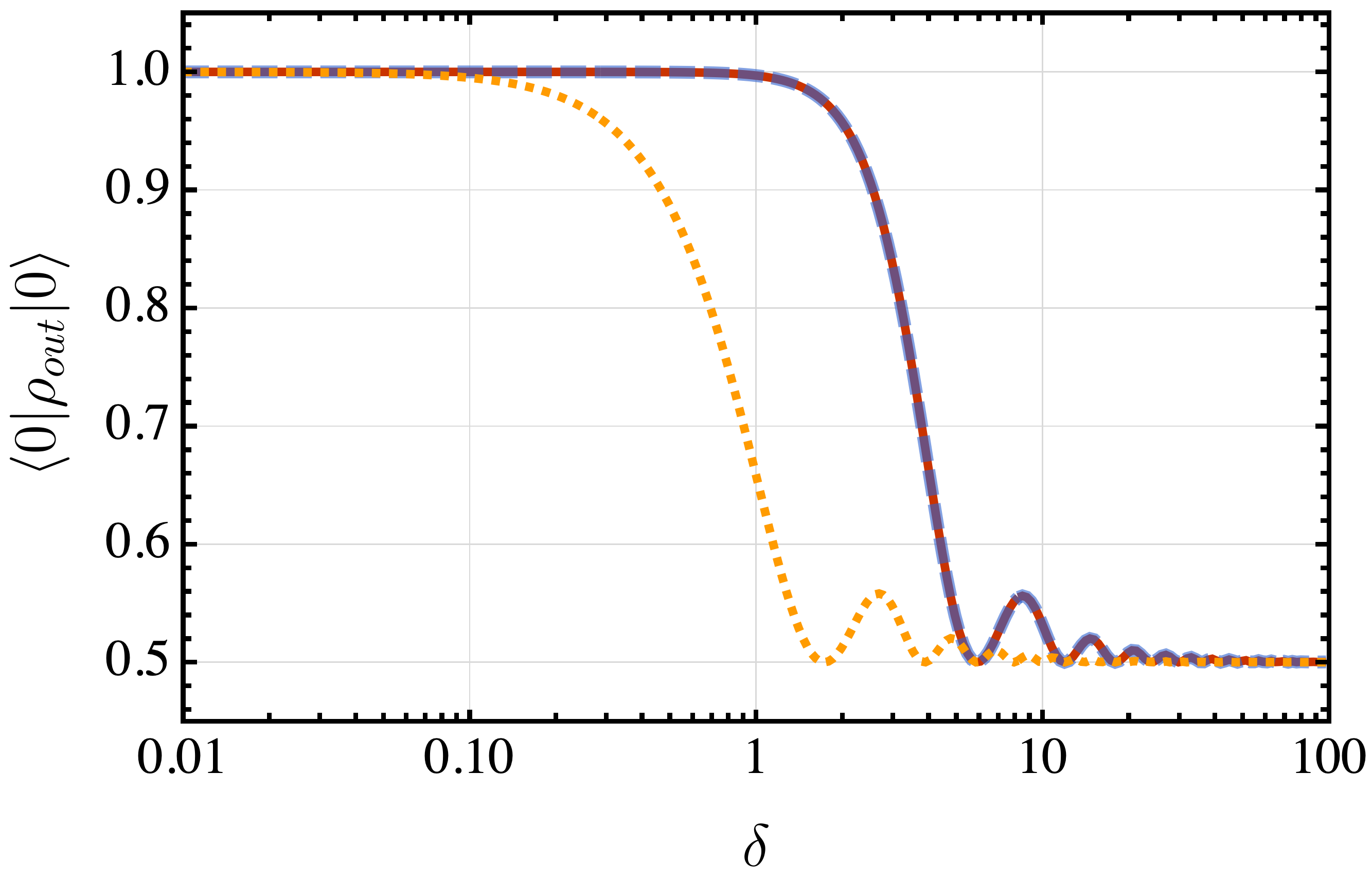}
  \caption{Comparison of population in $\ketbra{0}{0}$ of state $\rho_{\mathrm{out}}$ with respect to the dimensionless detuning $\delta = \Delta\omega/2J_{\perp}$, where $\rho_{\mathrm{out}}$ is either the result of not applying any operations (dotted orange) or applying independent unitary controls (dashed blue). The SDP benchmark is shown in solid red. When the controller is initially in the state $\rho_C=\ketbra{0}{0}$ and $\tau=\pi/2$, the set of independent unitary operations performs as well as the benchmark. The residual oscillations at large detuning arise from the finite interaction time and occur in a regime where population transfer is already strongly suppressed.}
  \label{fig:UnitaryVsFreeZeroState}
\end{figure}

First, we compare the free evolution dynamics of the target population with a restricted class of superinstruments corresponding to independent unitary operations acting locally on the controller at each of the two intervention times,
\begin{equation}
T = \Proj{\mathcal{U}_C(\vec{r})}\otimes \Proj{\mathcal{U}_C(\vec{s})},
\label{eq:UnitaryTester}
\end{equation}
where
\begin{equation}
U_C(\vec{r})=
\begin{pmatrix}
r_1 - \ii r_4 & -\ii r_2 - r_3 \\
-\ii r_2 + r_3 & r_1 + \ii r_4
\end{pmatrix},
\end{equation}
and the unitaries are parameterised by unit vectors in $\mathbb{R}^4$. The tensor product appears in \cref{eq:UnitaryTester} because the two unitary operations are independent and act on distinct input--output spaces (see \cref{eq:linkprod}). Since the set of superinstruments of the form \cref{eq:UnitaryTester} is non-convex, we write the corresponding optimisation problem as
\begin{equation}
\begin{aligned}
\max_{\vec{r},\vec{s}} \quad &
\matrixel{0}{
\tfrac12\id\otimes\rho_C\star W \star
\Proj{\mathcal{U}_C(\vec{r})}\otimes
\Proj{\mathcal{U}_C(\vec{s})}
}{0} \\
\text{s.t.}\quad &
\norm{\vec{r}}=1,\quad \norm{\vec{s}}=1.
\end{aligned}
\label{eq:UnitaryOptimisation}
\end{equation}
Although the above problem is guaranteed to admit a solution, the optimisation is non-convex and may therefore contain local optima. The SDP optimisation provides an upper bound against which this heuristic unitary search can be benchmarked. In the present two-step setting, the SDP in \cref{eq:sdp_indirect_control} takes the explicit form
\begin{equation}
\begin{aligned}
\max_T \quad &
\matrixel{0}{\tfrac12\id\otimes\rho_C\star W \star T}{0} \\
\text{s.t.}\quad &
\begin{aligned}
T &\geq 0, \\
\mathcal{L}_1(T) &= 0, \\
\mathcal{L}_2(T) &= 0,\\
\Tr(T) &= 4.
\end{aligned}
\end{aligned}
\label{eq:ResultsSDP}
\end{equation}
In \cref{fig:UnitaryVsFreeZeroState}, we compare the free evolution, the optimised independent-unitary strategy in \cref{eq:UnitaryOptimisation}, and the SDP benchmark in \cref{eq:ResultsSDP}. We initialise the controller in the state $\rho_C=\ketbra{0}{0}$ and fix the total interaction time to $J\tau=\pi/2$. Since we work in units of the coupling strength with $J=1$, this corresponds to $\tau=\pi/2$ and $t_0=t_1=t_2=\pi/6$. We then vary the detuning through the dimensionless parameter $\delta=\Delta\omega/(2J)$, shown on a logarithmic scale to capture the behaviour from the near-resonant to the far-detuned regime.

For small detuning, the free evolution already achieves almost complete polarisation of the target and there is consequently little room for improvement through control. As the detuning is increased, the performance of the free evolution decreases as population exchange between the system and controller becomes suppressed. The optimised unitary interventions, however, maintain near perfect polarisation over a substantially larger range of detunings. We also observe that for this choice of initial controller state, the optimised unitary strategy saturates the SDP benchmark to numerical precision over the entire range considered. Thus, no additional control resources provide an advantage for this particular initial state and interaction regime.

\begin{figure}[h]
  \centering
  \includegraphics[width=\columnwidth]{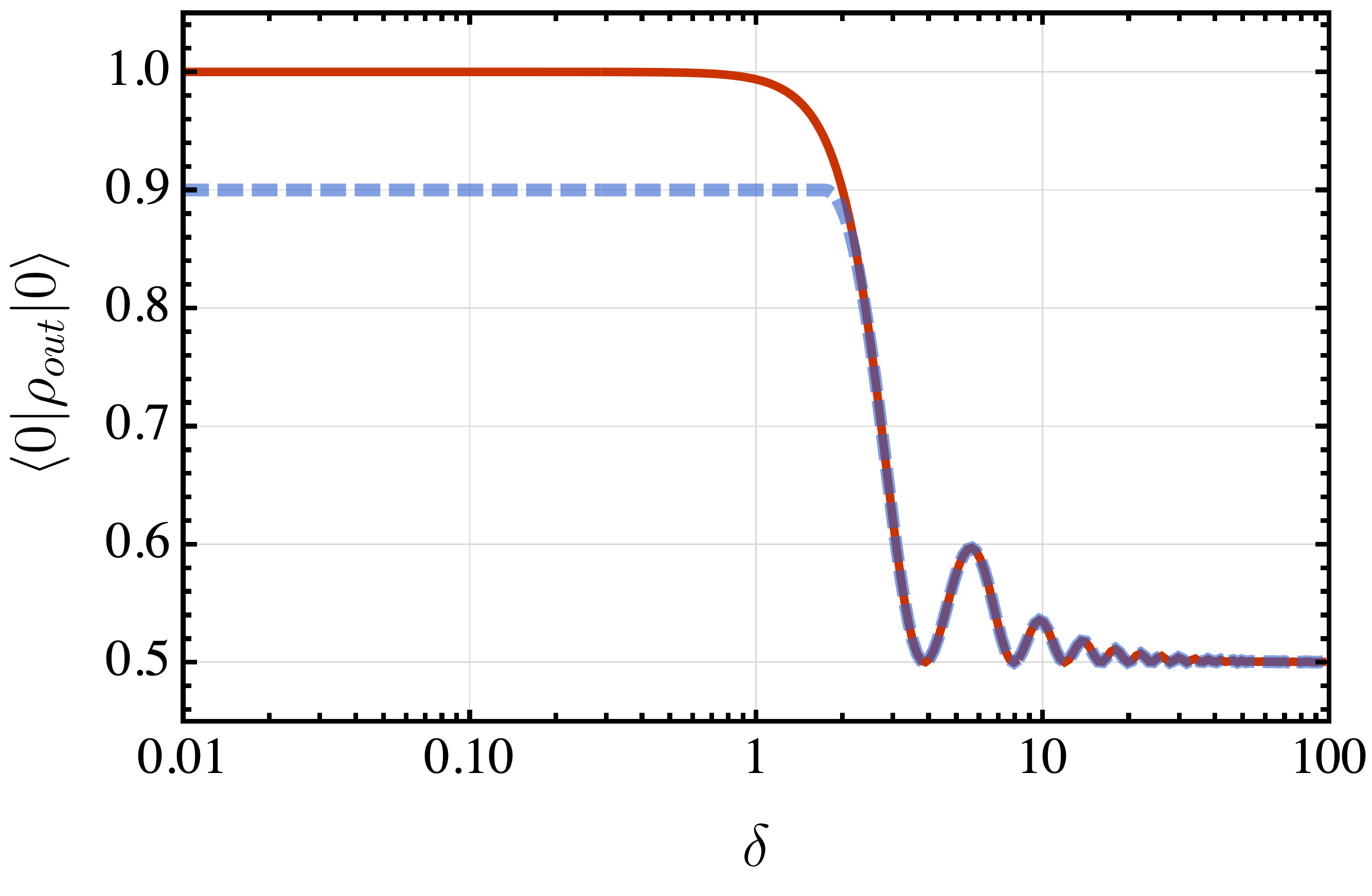}
  \caption{Comparison of population in $\ketbra{0}{0}$ of state $\rho_{\mathrm{out}}$ with respect to the dimensionless coupling $\delta = \Delta\omega/2J_{\perp}$, where $\rho_{\mathrm{out}}$ is either the result of applying the optimal general deterministic superinstrument (solid red) or uncorrelated unitaries (dashed blue). The controller is initially in the state $\rho_C=\frac{9}{10}\ketbra{0}{0}+\frac{1}{10}\ketbra{1}{1}$ and $\tau=3\pi/4$.}
  \label{fig:OptimalVsUnitaryMixed}
\end{figure}

However, when the controller is imperfectly initialised, unitary control is no longer sufficient to attain the optimal performance. To illustrate this, in \cref{fig:OptimalVsUnitaryMixed} we take $\rho_C=(1-\varepsilon)\ketbra{0}{0}+\varepsilon\ketbra{1}{1}$, with $\varepsilon=0.1$, and compare the optimised independent-unitary strategy from \cref{eq:UnitaryOptimisation} against the deterministic-superinstrument benchmark from \cref{eq:ResultsSDP}. In the low-detuning regime, the SDP continues to achieve almost perfect polarisation, whereas the unitary strategy is capped at $1-\varepsilon$. This limitation follows from the fact that the overall evolution generated by the fixed system--controller interactions together with local unitary controls is unitary on the joint system--controller state and therefore preserves its spectrum. For the initial state $\rho_S=\id/2$ and the mixed controller state above, the two largest eigenvalues of the joint state sum to $1-\varepsilon$, which gives the largest population that any unitary protocol can concentrate into the target subspace $\ketbra{0}{0}\otimes\mathcal{H}_C$ (See Appendix~\ref{app:free_evolution}). The observed gap from the SDP benchmark therefore reflects a genuine limitation of the unitary ansatz, rather than a failure of the heuristic optimisation, and demonstrates a regime in which more general control resources are necessary.

The results in \cref{fig:OptimalVsUnitaryMixed} show that independent unitary controls can be fundamentally insufficient when the controller is initially mixed. We therefore enlarge the admissible class of memoryless control operations by allowing a general quantum channel to act on the controller at each intervention time. Restricting again to independent operations at the two time steps, the corresponding superinstrument takes the form
\begin{equation}
  T=\Proj{\mathcal{M}_1}\otimes\Proj{\mathcal{M}_2},
  \label{eq:MarkovianTester}
\end{equation}
where $\Proj{\mathcal{M}_1}$ and $\Proj{\mathcal{M}_2}$ are the Choi operators of quantum channels acting on the controller. We refer to superinstruments of this form as Markovian superinstruments. We consider two heuristic approaches for constructing Markovian superinstruments. First, starting from the optimal deterministic superinstrument $T_{*}$ obtained from \cref{eq:ResultsSDP}, we construct a product superinstrument from its single-step marginals,
\begin{equation}
T = \frac14 \Tr_{C_I^{(2)}C_O^{(2)}}(T_{*})
\otimes
\Tr_{C_I^{(1)}C_O^{(1)}}(T_{*}).
\end{equation}
When relative entropy is used as the distance measure, this construction gives the Markovian superinstrument closest to $T_{*}$ in the corresponding sense~\cite{White_2022,Pollock2018}. However, the Markovian superinstrument closest to $T_{*}$ need not be the one that maximises the polarisation objective. As an alternative, we perform a see-saw optimisation directly over the two channels. We fix $\mathcal{M}_1$ and optimise $\mathcal{M}_2$ via an SDP, then fix $\mathcal{M}_2$ and optimise $\mathcal{M}_1$, iterating this procedure until convergence within numerical tolerance. Since the joint optimisation over $\mathcal{M}_1\otimes\mathcal{M}_2$ is non-convex, this procedure is only heuristic and is not guaranteed to converge to the global optimum over Markovian superinstruments.

\begin{figure}[h]
  \centering
  \includegraphics[width=\columnwidth]{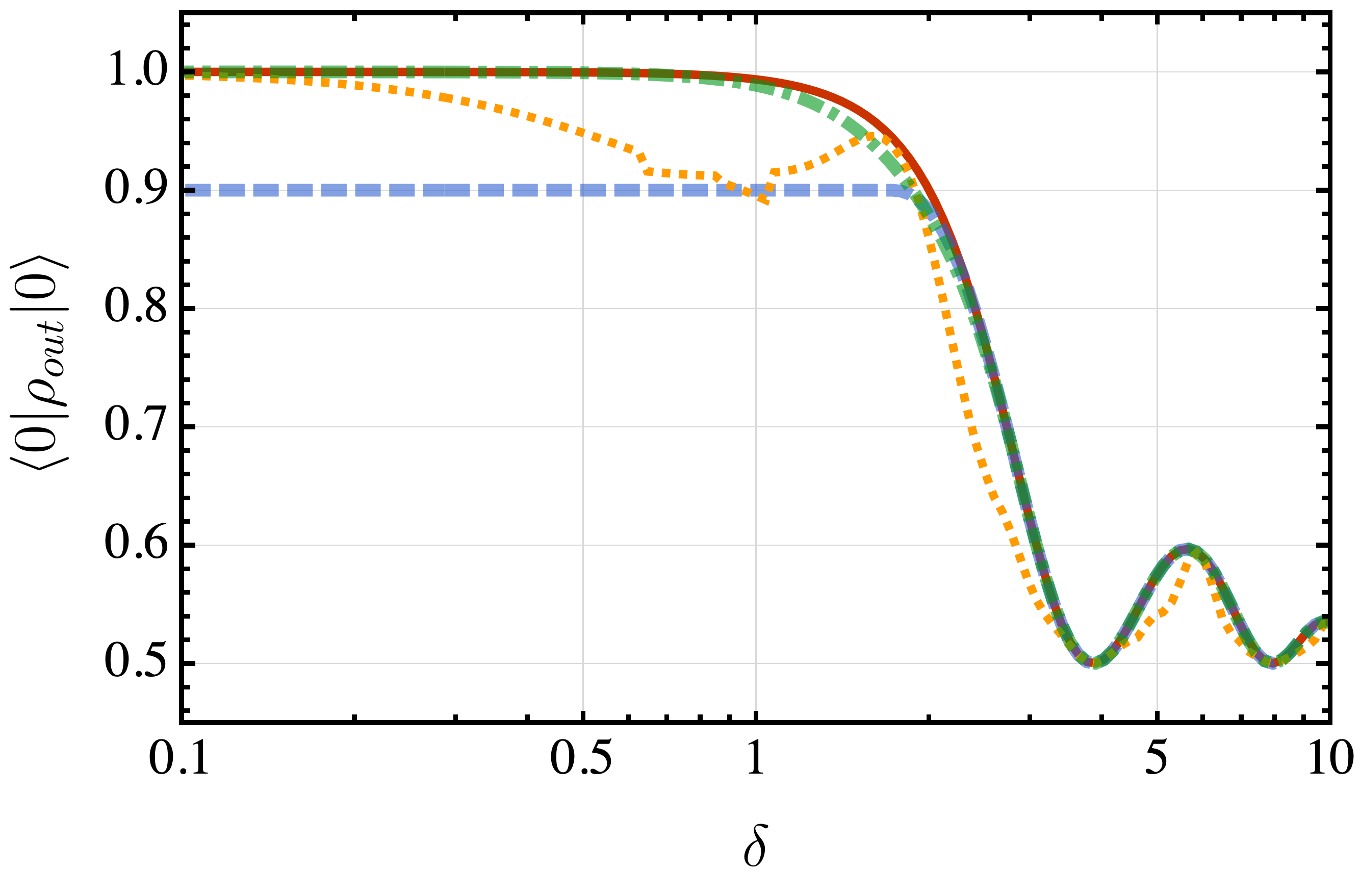}
  \caption{Population in $\ketbra{0}{0}$ of state $\rho_{\mathrm{out}}$ with respect to the dimensionless coupling $\delta = \Delta\omega/2J_{\perp}$ obtained by various heuristic methods benchmarked against the SDP over general deterministic superinstruments (solid red). The heuristic methods are optimisation over uncorrelated unitaries (dashed blue), see-saw optimisation over Markovian channels (dotted orange), and the Markovian channel closest to the optimal deterministic superinstrument (dashed-dotted green). The controller is initially in the state $\rho_C=\frac{9}{10}\ketbra{0}{0}+\frac{1}{10}\ketbra{1}{1}$.}
  \label{fig:HeuristicComparison}
\end{figure}

The performance of these two approaches is compared in \cref{fig:HeuristicComparison}. Allowing general channels can improve upon the unitary strategy, but the two Markovian constructions perform differently across the detuning range. In some regimes, the see-saw strategy outperforms the Markovian superinstrument obtained from the marginals of $T_{*}$, illustrating that proximity to the globally optimal superinstrument does not necessarily imply optimal task performance within the restricted Markovian class. Conversely, any remaining gap between either heuristic and the deterministic-superinstrument benchmark cannot by itself be attributed to the necessity of temporal correlations or feed-forward, since neither method certifies the global optimum over all Markovian superinstruments.
Although the example considered here involves only two control interventions on a single controller qubit, the resulting optimisation landscape is already highly non-linear and can be difficult to explore reliably with restricted ans\"atze. This highlights the usefulness of the SDP formulation as a benchmark. While the optimal deterministic superinstrument may require resources that are impractical or unnecessary in a given implementation, its performance provides a globally optimal reference against which simpler strategies can be assessed. In this way, the SDP allows us to distinguish genuine limitations of a restricted control class from limitations of the heuristic optimisation used to search within that class.

\section{Discussion and Conclusion}\label{sec:Conclusion}

In this work, we formulated the problem of indirect quantum control within the framework of higher-order quantum operations (HOQOs), modelling the most general causal control protocols by deterministic superinstruments. This class encompasses memoryless strategies, including independent unitaries, quantum channels, measurements, and state preparations applied at individual time steps, as well as more general temporally correlated protocols involving ancillary systems and information feed-forward. By choosing a suitable figure of merit for the final target state, the indirect-control problem can then be cast as an optimisation over deterministic superinstruments. When the figure of merit is linear in the final state, this optimisation takes the form of a semidefinite program (SDP), as shown in \cref{eq:sdp_indirect_control}. The SDP therefore provides an efficient way to optimise over the full convex set of admissible causal control protocols with a guarantee of convergence to the global optimum.

The deterministic superinstrument returned by the SDP represents an optimal protocol within this admissible causal class. In practice, however, its principal role may be as a benchmark rather than as a directly implementable control prescription. A generic optimal superinstrument can be nontrivial to decompose into experimentally available operations, and it may employ resources that are costly or undesirable in a particular implementation, such as coherent ancillary systems carrying quantum information between intervention times. Nevertheless, precisely because the SDP optimises over the full admissible class, its optimal value provides a rigorous performance benchmark. More restricted and experimentally motivated control strategies can then be assessed by comparing their achievable performance against this optimum.

The two-step polarisation example illustrates how the benchmark can be used to assess the value of different control resources. When the controller is initially prepared in the pure state $\rho_C=\ketbra{0}{0}$, we find that optimised independent unitary controls saturate the deterministic-superinstrument benchmark over the full detuning range considered, showing that more general resources provide no additional advantage in this regime. This changes when the controller is imperfectly initialised, where the unitary strategy is fundamentally limited while more general control operations can achieve higher polarisation. Allowing independent quantum channels can recover part of this performance gap, although the corresponding optimisation remains non-convex and the heuristic methods considered here do not certify the global optimum over all Markovian superinstruments.

These observations also point to several natural directions for future work. In particular, when the Markovian heuristics fail to attain the deterministic-superinstrument benchmark, the present analysis does not establish whether the remaining gap is due to the suboptimality of the heuristic search or whether genuinely temporally correlated resources, such as classical or quantum feed-forward, are required. A useful extension of the framework would therefore be to develop a systematic hierarchy of admissible control resources, ranging from independent unitaries and general memoryless channels to protocols incorporating classical memory and coherent quantum memory across intervention times. Such a hierarchy would allow one to identify more precisely when temporal correlations constitute a genuine resource for indirect control. A further challenge is scalability since the dimension of the process tensor and superinstrument grows rapidly with the number of intervention times. Tensor-network representations, such as matrix-product-operator descriptions of multi-time processes, may provide a route towards extending these benchmarks to longer control sequences, as recently explored in the context of non-Markovian quantum control~\cite{Ortega_2024}.

We believe that the analysis in \cref{sec:Results} can help inform the design of short indirect-control sequences by providing a rigorous performance benchmark together with an optimal protocol when arbitrary admissible resources are allowed. The polarisation task considered here may also be viewed as a simple instance of indirect cooling \cite{Burgarth2007}, in which an initially hot, maximally mixed qubit is driven towards its ground state through control of an auxiliary system. A natural extension would be to replace the target qubit by a harmonic oscillator and investigate whether similar oscillator--qubit interactions can be exploited for indirect cooling. More broadly, the framework provides a systematic way to separate what is achievable in principle from what can be achieved with experimentally restricted control resources, and thereby to quantify the value of increasingly general control strategies.

\begin{acknowledgments}
We acknowledge the role of Sarath Raman Nair for the original project idea and ongoing feedback and discussions. \v{S}V, VS, and RG acknowledge the support from the Sydney Quantum Academy, Sydney, Australia. VS also acknowledges funding support by the Hon-Hai Research Institute through the Australian Quantum Software Network (AQSN) microgrant.
\end{acknowledgments}

\appendix
\crefalias{section}{appendix}

\section{Proof of \texorpdfstring{\Cref{eq:final_output}}{(eq:final_output)}}
\label{app:Linkprod}

Given two maps $\mathcal{M} : \mathcal{L}(\mathcal{H}^{A}) \rightarrow \mathcal{L}(\mathcal{H}^{B})$ and $\mathcal{N} : \mathcal{L}(\mathcal{H}^{B}) \rightarrow \mathcal{L}(\mathcal{H}^{C})$, the transformation of a state $\rho \in \mathcal{L}(\mathcal{H}^A)$ under the composition $\rho_{f}=\mathcal{N} \circ \mathcal{M} (\rho)$ can be represented in terms of the Choi operators as
\begin{equation}
  \label{eq:link_append_example}
  \Proj{N}^{BC} \star \Proj{M}^{AB} \star \rho^{A}.
\end{equation}
Here the superscripts label the corresponding Hilbert spaces. As shown in the definition of the link product in \cref{eq:linkprod}, the operation contracts the common Hilbert spaces, up to a partial transpose. Therefore, once the Hilbert-space labels are tracked explicitly, the link product provides a compact notation for composing quantum maps in their Choi representation.

The link product has several useful properties. For operators $A$, $B$, and $C$, it is Hermiticity preserving, i.e., if $A$ and $B$ are Hermitian, then $A\star B$ is Hermitian. It is also positivity preserving, so that $A,B\geq 0$ implies $A\star B\geq 0$. Moreover, the link product is commutative,
$A\star B = B\star A$, and associative,
$A\star (B\star C) = (A\star B)\star C$,
provided the Hilbert-space labels are consistently specified. Therefore in \cref{eq:link_append_example}, the individual terms can be written in arbitrary order by keeping track of the specific Hilbert space labels.

Moreover, if $A$ and $B$ are operators on the same Hilbert space, then
\begin{equation}
  \label{eq:ApeendA:transposelink}
A \star B= \Tr\left(A^{T} B\right).
\end{equation}
On the other hand, if $A$ and $B$ do not share any common Hilbert-space factor, then
\begin{equation}
  \label{eq:ApeendA:tensorprodlink}
A \star B= A \otimes B.
\end{equation}

Given the indirect-control setup in \cref{sec:HOQO}, the output state can be written as
\begin{equation}
  \label{eq:Append_rho_out}
  \rho_{\mathrm{out}}=\Tr_{AC}\left(\mathcal{U}_{SC}^{(N)}\circ \ldots\mathcal{U}_{CA}^{(2)}\circ\mathcal{U}_{SC}^{(1)}\circ(\ketbra{\psi_{ini}}{\psi_{ini}})\right)
\end{equation}
where
$\ketbra{\psi_{ini}}{\psi_{ini}}=\ketbra{\psi_{S}}{\psi_{S}} \otimes \ketbra{\psi_{C}}{\psi_{C}} \otimes \ketbra{\psi_{A}}{\psi_{A}}$.
Following the same Choi-representation argument used in \cref{eq:link_append_example}, this sequential composition can be rewritten using the link product. The commutativity of the link product then allows us to group the fixed parts of the dynamics, namely the $S$--$C$ interactions $\mathcal{U}_{SC}^{(t)}$, separately from the controllable operations $\mathcal{U}_{CA}^{(t)}$. Using \cref{eq:ApeendA:transposelink,eq:ApeendA:tensorprodlink}, this gives the decomposition stated in \cref{eq:final_output,eq:process_tensor,eq:superinstrument_def}.

\section{Proof of optimality of solution of SDP}
\label{app:DualSDP}

In this appendix we derive the dual of the SDP in \cref{eq:sdp_indirect_control}. Any dual-feasible point provides an upper bound on the performance of every admissible deterministic superinstrument. Showing that strong duality holds then certifies that the optimum of \cref{eq:sdp_indirect_control} is the optimal benchmark over this class.

We define
\begin{equation}
G
=
\rho_S
\otimes
\rho_C
\star W \star \Omega^{T},
\end{equation}
so that
\begin{equation}
  \Tr\!\left[
    \Omega\!\left(
      \rho_S
      \otimes
      \rho_C
      \star W \star T
    \right)
  \right]
  =
  \Tr\!\left(GT\right).
\end{equation}
where the transpose in the definition of \(G\) is determined by the Choi and link-product convention used in \cref{eq:linkprod}. With this notation, the primal SDP is

\begin{equation}
  \label{eq:primal_sdp_appendix}
  \begin{aligned}
  \max_T \quad
  & \Tr\!\left(GT\right)\\
  \text{s.t.}\quad
  & T \geq 0,\\
  & \mathcal{L}_{n}(T)=0,
  \qquad n=1,\ldots,N,\\
  & \Tr\!\left(T\right)=D_I .
  \end{aligned}
\end{equation}

and we can introduce a Lagrangian
\begin{equation}
\begin{aligned}
\mathbb{L}
=&\;
\Tr\!\left(GT\right)
+
\Tr\!\left(ZT\right)
+
\lambda\left(D_I-\Tr\!\left(T\right)\right) \\
&+
\sum_{n=1}^{N}
\Tr\!\left(Y_n \mathcal{L}_{n}(T)\right),
\end{aligned}
\end{equation}
where \(Z\geq 0\), \(Y_n=Y_n^\dagger\), and \(\lambda\in\mathbb{R}\). The maps \(\mathcal{L}_n\) are self-adjoint with respect to the Hilbert--Schmidt inner product, so that

\begin{equation}
\Tr\!\left(Y_n \mathcal{L}_{n}(T)\right)
=
\Tr\!\left(\mathcal{L}_{n}(Y_n)T\right).
\end{equation}

Hence the Lagrangian can be rearranged as
\begin{equation}
\mathbb{L}
=
\lambda D_I
+
\Tr\!\left[
\left(
G+Z-\lambda\mathbb{I}
+
\sum_{n=1}^{N}\mathcal{L}_{n}(Y_n)
\right)T
\right].
\end{equation}
Minimising the Lagrangian therefore gives us the condition
\begin{equation}
Z
=
\lambda\mathbb{I}
-
\sum_{n=1}^{N}\mathcal{L}_{n}(Y_n)
-
G .
\end{equation}
Since \(Z\geq0\), the dual SDP is
\begin{equation}
  \label{eq:dualsdp_indirect_control}
  \begin{aligned}
  \min_{\lambda,\{Y_n\}} \quad
  & \lambda D_I \\
  \text{s.t.}\quad
  &
  \lambda\mathbb{I}
  -
  \sum_{n=1}^{N}\mathcal{L}_{n}(Y_n)
  -
  G
  \geq 0,\\
  &Y_n=Y_n^\dagger,
  \qquad n=1,\ldots,N,\\
  &\lambda\in\mathbb{R}.
  \end{aligned}
\end{equation}

The upper-bound property of the dual follows directly. Let \(T\) be any feasible deterministic superinstrument. For any dual-feasible \(\lambda,\{Y_n\},Z\),

\begin{align}
\lambda D_I-\Tr\!\left(GT\right)
&=
\Tr\!\left(\lambda\mathbb{I}T\right)
-
\Tr\!\left(GT\right) \\
&=
\Tr\!\left[
\left(
\lambda\mathbb{I}
-
G
\right)T
\right] \\
&=
\Tr\!\left[
\left(
Z+
\sum_{n=1}^{N}\mathcal{L}_{n}(Y_n)
\right)T
\right] \\
&=
\Tr\!\left(ZT\right)
+
\sum_{n=1}^{N}
\Tr\!\left(Y_n\mathcal{L}_{n}(T)\right).
\end{align}
Using the primal feasibility conditions \(\mathcal{L}_{n}(T)=0\), we obtain

\begin{equation}
\lambda D_I-\Tr\!\left(GT\right)
=
\Tr\!\left(ZT\right).
\end{equation}
Since \(Z\geq0\) and \(T\geq0\), it follows that
\begin{equation}
\Tr\!\left(GT\right)\leq \lambda D_I .
\end{equation}
Thus every dual-feasible point gives an upper bound on the performance of every admissible deterministic superinstrument. Strong duality holds as both primal and dual SDPs are strictly feasible \cite{Skrzypczyk_23}. A strictly feasible primal point is given, for example, by applying completely depolarising channels at each slot. A strictly feasible dual point is obtained by taking \(Y_n=0\) and choosing \(\lambda\) sufficiently large so that \(\lambda\mathbb{I}-G>0\). Hence the optimal primal and dual values coincide.

Let \(T^{*}\) be an optimal deterministic superinstrument and let
\(\lambda^{*},Y_n^{*},Z^{*}\) be an optimal dual solution. Strong
duality gives
\begin{equation}
\Tr\!\left(GT^{*}\right)
=
\lambda^{*}D_I .
\end{equation}

Hence the solution of \cref{eq:sdp_indirect_control} provides the optimal benchmark value over the full class of deterministic control protocols.

\section{Derivation of the free-evolution dynamics and maximum efficiency under imperfect controller initialisation}
\label{app:free_evolution}

Here we derive the free-evolution expression in \cref{eq:free_evol}. The system--controller Hamiltonian is

\begin{equation}
\begin{aligned}
H &= \frac{\omega}{2}\sigma_z^{(S)}
   + \frac{\omega+\Delta\omega}{2}\sigma_z^{(C)} \\
  &\quad + J_{\parallel}\sigma_z^{(S)}\sigma_z^{(C)} \\
  &\quad + J_{\perp}\left(
    \sigma_+^{(S)}\sigma_-^{(C)}
    + \sigma_-^{(S)}\sigma_+^{(C)}
    \right).
\end{aligned}
\label{eq:app_hamiltonian}
\end{equation}

The transverse interaction conserves the total number of excitations and only couples the states $\ket{01}$ and $\ket{10}$. Consequently, the joint Hilbert space decomposes into the two invariant subspaces
\begin{equation}
\mathcal{H}_{\mathrm{e}}
=
\operatorname{span}\{\ket{00},\ket{11}\},
\quad
\mathcal{H}_{\mathrm{o}}
=
\operatorname{span}\{\ket{01},\ket{10}\}.
\end{equation}
In the ordered basis $\{\ket{00},\ket{11}\}$, the Hamiltonian is diagonal,
\begin{equation}
H_{\mathrm{e}}
=
\begin{pmatrix}
\omega+\frac{\Delta\omega}{2}+J_{\parallel} & 0 \\[2mm]
0 & -\omega-\frac{\Delta\omega}{2}+J_{\parallel}
\end{pmatrix},
\label{eq:app_even_sector}
\end{equation}
so neither $\ket{00}$ nor $\ket{11}$ undergoes population transfer.

The non-trivial population dynamics takes place in the single-excitation subspace $\mathcal{H}_{\mathrm{o}}$. In the ordered basis $\{\ket{01},\ket{10}\}$, the corresponding block of the Hamiltonian is
\begin{equation}
H_{\mathrm{o}}
=
-J_{\parallel}\id
+
\begin{pmatrix}
-\frac{\Delta\omega}{2} & J_{\perp} \\[2mm]
J_{\perp} & \frac{\Delta\omega}{2}
\end{pmatrix}.
\label{eq:app_odd_sector}
\end{equation}
The common frequency $\omega$ therefore drops out of this sector, while the longitudinal interaction $J_{\parallel}$ contributes only a term proportional to the identity. It consequently produces an overall phase and does not affect the population exchange between $\ket{01}$ and $\ket{10}$ during free evolution. Defining
\begin{equation}
\Omega
=
\sqrt{
J_{\perp}^{2}
+
\frac{\Delta\omega^{2}}{4}
},
\label{eq:app_omega_eff}
\end{equation}
Using the convention $U_{SC}(\tau)=\exp(\ii\tau H)$ adopted
in the main text, its action on $\ket{10}$ is therefore
\begin{align}
U_{SC}(\tau)\ket{10}
={}&
e^{-\ii J_{\parallel}\tau}
\Bigg[
\left(
\cos(\Omega\tau)
+
\ii\frac{\Delta\omega}{2\Omega}
\sin(\Omega\tau)
\right)\ket{10}
\nonumber\\
&\hspace{2.8cm}
+
\ii\frac{J_{\perp}}{\Omega}
\sin(\Omega\tau)\ket{01}
\Bigg].
\label{eq:app_state_evolution}
\end{align}
It follows that the probability for population exchange from $\ket{10}$ to $\ket{01}$ is
\begin{equation}
P_{10\rightarrow01}(\tau)
=
\frac{J_{\perp}^{2}}{\Omega^{2}}
\sin^{2}(\Omega\tau)
\label{eq:app_exchange_probability}
\end{equation}

For the initial states considered in the main text,
\begin{equation}
\rho_S=\frac{1}{2}\id,
\qquad
\rho_C=\ketbra{0}{0},
\end{equation}
the initial joint state is
\begin{equation}
\rho_{SC}(0)
=
\frac{1}{2}\ketbra{00}{00}
+
\frac{1}{2}\ketbra{10}{10}
\label{eq:app_initial_state}
\end{equation}
The first component, $\ket{00}$, is an eigenstate of the Hamiltonian and hence always contributes $1/2$ to the final population of $\ket{0}$ in the target system. For the second component, initially $\ket{10}$, the target occupies $\ket{0}$ only when population is transferred to $\ket{01}$. Consequently,
\begin{equation}
f(\mathcal{I})
=
\frac{1}{2}
+
\frac{1}{2}P_{10\rightarrow01}(\tau).
\label{eq:app_free_intermediate}
\end{equation}
Introducing the dimensionless detuning
\begin{equation}
\delta
=
\frac{\Delta\omega}{2J_{\perp}},
\end{equation}
such that
\begin{equation}
\Omega
=
J_{\perp}\sqrt{1+\delta^{2}},
\end{equation}
we finally obtain
\begin{equation}
f(\mathcal{I})
=
\frac{1}{2}
+
\frac{1}{2}
\left[
\frac{1}{1+\delta^{2}}
\sin^{2}
\left(
J_{\perp}\tau\sqrt{1+\delta^{2}}
\right)
\right],
\end{equation}
which is \cref{eq:free_evol}.

For an imperfectly initialised controller, $\rho_C=(1-\varepsilon)\ketbra{0}{0}+\varepsilon\ketbra{1}{1}$ with $0\leq\varepsilon\leq1/2$, the initial state $\rho_{SC}=\id_S/2\otimes\rho_C$ satisfies $\rho_{SC}\leq(1-\varepsilon)\id_{SC}/2$. Hence, for every joint unitary $U$ on $S\otimes C$ and the rank-two projector $P=\ketbra{0}{0}_S\otimes\id_C$,
\begin{equation}
\operatorname{Tr}\!\left(PU\rho_{SC}U^\dagger\right)
\leq \frac{1-\varepsilon}{2}\operatorname{Tr}P
=1-\varepsilon.
\label{eq:app_unitary_polarization_bound}
\end{equation}
The bound is attained by SWAP when arbitrary joint unitaries are allowed. Every sequence of fixed system--controller interactions and local unitary controls composes to a joint unitary, so the same bound applies to all such sequences. Therefore, any achievable deterministic-superinstrument value above $1-\varepsilon$ certifies a strict advantage over all unitary protocols on $SC$, independently of the heuristic optimisation.

\bibliography{references}% Produces the bibliography via BibTeX.

\end{document}

%% file: Figures/Tikz/process-tester-comb.tex
\begingroup
\tikzset{
  process-blob/.style={
    rounded corners,
    fill=yellow!30,
    inner xsep=2pt,
    draw=none,
    yshift=2pt
  },
  process-label/.style={
    label position=above,
    anchor=north,
    yshift=0.4cm
  },
  tester-blob/.style={
    rounded corners,
    fill=blue!10,
    inner xsep=2pt,
    draw=none,
    yshift=-2pt
  },
  tester-label/.style={
    label position=below,
    anchor=north,
    yshift=-0.2cm
  }
}

\begin{quantikz}[
  column sep=0.75cm
]
  \rho_{S}\,
  & \gate[2]{\mathcal{U}_{SC}^{(0)}}
  \gategroup[2,steps=1,
    style={process-blob},
    background,
    label style={process-label}
  ]{}
  \gategroup[1,steps=10,
    style={process-blob},
    background,
    label style={process-label}
  ]{$W$}
  && \gate[2]{\mathcal{U}_{SC}^{(1)}}
  \gategroup[2,steps=1,
    style={process-blob},
    background,
    label style={process-label}
  ]{}
  && \gate[2]{\mathcal{U}_{SC}^{(2)}}
  \gategroup[2,steps=1,
    style={process-blob},
    background,
    label style={process-label}
  ]{}
  &&  \ldots \ldots &&
  & \gate[2]{\mathcal{U}_{SC}^{(N)}}
  \gategroup[2, steps=1,
    style={process-blob},
    background,
    label style={process-label}
  ]{}
  & \,\rho_{\mathrm{out}}
\\
  \rho_{C}\,
  & \wire[l][1]["C_O^{(0)}"{above,pos=0.5}]{a}
  & \gate[2]{\mathcal{U}_{CA}^{(1)}} \wire[l][1]["C_I^{(1)}"{above,pos=0.5}]{a}
  \gategroup[2,steps=1,
    style={tester-blob},
    background,
    label style={tester-label}
  ]{}
  & \wire[l][1]["C_O^{(1)}"{above,pos=0.5}]{a}
  & \gate[2]{\mathcal{U}_{CA}^{(2)}} \wire[l][1]["C_I^{(2)}"{above,pos=0.5}]{a}
  \gategroup[2,steps=1,
    style={tester-blob},
    background,
    label style={tester-label}
  ]{}
  & \wire[l][1]["C_O^{(2)}"{above,pos=0.5}]{a}
  & \wire[l][1]["C_I^{(3)}"{above,pos=0.5}]{a}
  & \ldots \ldots &
  & \gate[2]{\mathcal{U}_{CA}^{(N)}} \wire[l][1]["C_I^{(N)}"{above,pos=0.5}]{a}
  \gategroup[2,steps=1,
    style={tester-blob},
    background,
    label style={tester-label}
  ]{}
  & \wire[l][1]["C_O^{(N)}"{above,pos=0.5}]{a}
  \gategroup[1,steps=3,
    style={process-blob},
    background,
    label style={process-label}
  ]{}
  & \wire[l][1]["C_I^{(N+1)}"{above, xshift=0.2cm, pos=0.5}]{a}
  & \slash{}
\\
  \rho_{A}\
  \gategroup[1,steps=12,
    style={tester-blob},
    background,
    label style={tester-label}
  ]{$T$}
  &&&&&&&  \ldots \ldots &&&
  & \slash{} & \setwiretype{n}
\end{quantikz} 

\endgroup

%% file: Figures/Tikz/purification-diagram.tex
\begingroup
\tikzset{
  process-blob/.style={
    rounded corners,
    fill=yellow!30,
    inner xsep=2pt,
    draw=none,
    yshift=2pt
  },
  process-label/.style={
    label position=above,
    anchor=north,
    yshift=0.2cm
  },
  tester-blob/.style={
    rounded corners,
    fill=blue!10,
    inner xsep=2pt,
    draw=none,
    yshift=-2pt
  },
  tester-label/.style={
    label position=below,
    anchor=north,
    yshift=-0.2cm
  }
}

\begin{quantikz}
  \rho_S\,
  & \gate[2]{\mathcal{U}_{SC}^{(0)}}
  \gategroup[2,steps=1,
    style={process-blob},
    background,
    label style={process-label}
  ]{}
  \gategroup[1,steps=5,
    style={process-blob},
    background,
    label style={process-label}
  ]{$W$}
  && \gate[2]{\mathcal{U}_{SC}^{(1)}}
  \gategroup[2,steps=1,
    style={process-blob},
    background,
    label style={process-label}
  ]{}
  && \gate[2]{\mathcal{U}_{SC}^{(2)}}
  \gategroup[2,steps=1,
    style={process-blob},
    background,
    label style={process-label}
  ]{}
  & \,\rho_{\mathrm{out}}
\\
  \rho_C\,
  && \gate[1,style={draw=none,fill=none},label style={opacity=0}]{\hphantom{\mathcal{M}_{1}}}
  \gategroup[2,steps=1,
    style={tester-blob},
    background,
    label style={tester-label}]{}
  && \gate[1,style={draw=none,fill=none},label style={opacity=0}]{\hphantom{\mathcal{M}_{2}}}
  \gategroup[2,steps=1,
    style={tester-blob},
    background,
    label style={tester-label}
  ]{}
  & \gategroup[1,steps=2,
    style={process-blob},
    background,
    label style={tester-label}
  ]{}
  & \slash{}
\\
  \setwiretype{n}
  && \wireoverride{2}
  \gategroup[1,steps=3,
    style={tester-blob},
    background,
    label style={tester-label}
  ]{$T$}
  &&&&&
\end{quantikz} 

\endgroup

%% file: references.bib
@article{Warren1993,
author = {Warren S. Warren  and Herschel Rabitz  and Mohammed Dahleh },
title = {Coherent Control of Quantum Dynamics: The Dream Is Alive},
journal = {Science},
volume = {259},
number = {5101},
pages = {1581-1589},
year = {1993},
doi = {10.1126/science.259.5101.1581},
URL = {https://www.science.org/doi/abs/10.1126/science.259.5101.1581},
eprint = {https://www.science.org/doi/pdf/10.1126/science.259.5101.1581}}

@ARTICLE{Chu2002-ua,
  title    = "Cold atoms and quantum control",
  author   = "Chu, Steven",
  journal  = "Nature",
  volume   =  416,
  number   =  6877,
  pages    = "206--210",
  month    =  mar,
  year     =  2002
}

@book{blaquiere2014information,
  title={Information Complexity and Control in Quantum Physics: Proceedings of the 4th International Seminar on Mathematical Theory of Dynamical Systems and Microphysics Udine, September 4--13, 1985},
  author={Blaquiere, A. and Diner, S. and Lochak, G.},
  isbn={9783709129715},
  series={CISM International Centre for Mechanical Sciences},
  url={https://books.google.com.au/books?id=fD8sBAAAQBAJ},
  year={2014},
  publisher={Springer Vienna}
}

@book{wiseman2009quantum,
  title={Quantum Measurement and Control},
  author={Wiseman, H.M. and Milburn, G.J.},
  isbn={9781139482912},
  url={https://books.google.com.au/books?id=8H8hAwAAQBAJ},
  year={2009},
  publisher={Cambridge University Press}
}

@book{1983differential,
  title={Differential Geometric Control Theory: Proceedings, Michigan Technological University, 28.6.-2.7.82},
  isbn={9780817630911},
  series={Progress in Mathematics},
  url={https://books.google.com.au/books?id=szB-AAAACAAJ},
  year={1983},
  publisher={Birkh{\"a}user Boston}
}

@article{Burgarth2009,
  title = {Local controllability of quantum networks},
  author = {Burgarth, Daniel and Bose, Sougato and Bruder, Christoph and Giovannetti, Vittorio},
  journal = {Phys. Rev. A},
  volume = {79},
  issue = {6},
  pages = {060305},
  numpages = {4},
  year = {2009},
  month = {Jun},
  publisher = {American Physical Society},
  doi = {10.1103/PhysRevA.79.060305},
  url = {https://link.aps.org/doi/10.1103/PhysRevA.79.060305}
}

@article{Burgarth2010,
  title = {Scalable quantum computation via local control of only two qubits},
  author = {Burgarth, Daniel and Maruyama, Koji and Murphy, Michael and Montangero, Simone and Calarco, Tommaso and Nori, Franco and Plenio, Martin B.},
  journal = {Phys. Rev. A},
  volume = {81},
  issue = {4},
  pages = {040303},
  numpages = {4},
  year = {2010},
  month = {Apr},
  publisher = {American Physical Society},
  doi = {10.1103/PhysRevA.81.040303},
  url = {https://link.aps.org/doi/10.1103/PhysRevA.81.040303}
}

@ARTICLE{Morton2006,
  title    = "Bang--bang control of fullerene qubits using ultrafast phase
              gates",
  author   = "Morton, John J L and Tyryshkin, Alexei M and Ardavan, Arzhang and
              Benjamin, Simon C and Porfyrakis, Kyriakos and Lyon, S A and
              Briggs, G Andrew D",
  journal  = "Nature Physics",
  volume   =  2,
  number   =  1,
  pages    = "40--43",
  month    =  jan,
  year     =  2006
}

@article{Hodges2008,
  title = {Universal control of nuclear spins via anisotropic hyperfine interactions},
  author = {Hodges, J. S. and Yang, J. C. and Ramanathan, C. and Cory, D. G.},
  journal = {Phys. Rev. A},
  volume = {78},
  issue = {1},
  pages = {010303},
  numpages = {4},
  year = {2008},
  month = {Jul},
  publisher = {American Physical Society},
  doi = {10.1103/PhysRevA.78.010303},
  url = {https://link.aps.org/doi/10.1103/PhysRevA.78.010303}
}

@ARTICLE{Taminiau2014,
  title    = "Universal control and error correction in multi-qubit spin
              registers in diamond",
  author   = "Taminiau, T H and Cramer, J and van der Sar, T and Dobrovitski, V
              V and Hanson, R",
  journal  = "Nature Nanotechnology",
  volume   =  9,
  number   =  3,
  pages    = "171--176",
  month    =  mar,
  year     =  2014
}

@article{Peirce1988,
  title = {Optimal control of quantum-mechanical systems: Existence, numerical approximation, and applications},
  author = {Peirce, Anthony P. and Dahleh, Mohammed A. and Rabitz, Herschel},
  journal = {Phys. Rev. A},
  volume = {37},
  issue = {12},
  pages = {4950--4964},
  numpages = {0},
  year = {1988},
  month = {Jun},
  publisher = {American Physical Society},
  doi = {10.1103/PhysRevA.37.4950},
  url = {https://link.aps.org/doi/10.1103/PhysRevA.37.4950}
}

@article{Huang1983,
    author = {Huang, Garng M. and Tarn, T. J. and Clark, John W.},
    title = {On the controllability of quantum mechanical systems},
    journal = {Journal of Mathematical Physics},
    volume = {24},
    number = {11},
    pages = {2608-2618},
    year = {1983},
    month = {11},
    issn = {0022-2488},
    doi = {10.1063/1.525634},
    url = {https://doi.org/10.1063/1.525634},
}

@article{Cappellaro2009,
  title = {Coherence and Control of Quantum Registers Based on Electronic Spin in a Nuclear Spin Bath},
  author = {Cappellaro, P. and Jiang, L. and Hodges, J. S. and Lukin, M. D.},
  journal = {Phys. Rev. Lett.},
  volume = {102},
  issue = {21},
  pages = {210502},
  numpages = {4},
  year = {2009},
  month = {May},
  publisher = {American Physical Society},
  doi = {10.1103/PhysRevLett.102.210502},
  url = {https://link.aps.org/doi/10.1103/PhysRevLett.102.210502}
}

@article{Lloyd2001,
  title = {Engineering quantum dynamics},
  author = {Lloyd, Seth and Viola, Lorenza},
  journal = {Phys. Rev. A},
  volume = {65},
  issue = {1},
  pages = {010101},
  numpages = {4},
  year = {2001},
  month = {Dec},
  publisher = {American Physical Society},
  doi = {10.1103/PhysRevA.65.010101},
  url = {https://link.aps.org/doi/10.1103/PhysRevA.65.010101}
}

@article{Lloyd2004,
  title = {Universal quantum interfaces},
  author = {Lloyd, Seth and Landahl, Andrew J. and Slotine, Jean-Jacques E.},
  journal = {Phys. Rev. A},
  volume = {69},
  issue = {1},
  pages = {012305},
  numpages = {4},
  year = {2004},
  month = {Jan},
  publisher = {American Physical Society},
  doi = {10.1103/PhysRevA.69.012305},
  url = {https://link.aps.org/doi/10.1103/PhysRevA.69.012305}
}

@article{Burgarth2007,
  title = {Full Control by Locally Induced Relaxation},
  author = {Burgarth, Daniel and Giovannetti, Vittorio},
  journal = {Phys. Rev. Lett.},
  volume = {99},
  issue = {10},
  pages = {100501},
  numpages = {4},
  year = {2007},
  month = {Sep},
  publisher = {American Physical Society},
  doi = {10.1103/PhysRevLett.99.100501},
  url = {https://link.aps.org/doi/10.1103/PhysRevLett.99.100501}
}

@ARTICLE{DAlessandro2012,
  author={D'Alessandro, Domenico and Romano, Raffaele},
  journal={IEEE Transactions on Automatic Control}, 
  title={Indirect Controllability of Quantum Systems; A Study of Two Interacting Quantum Bits}, 
  year={2012},
  volume={57},
  number={8},
  pages={2009-2020},
  doi={10.1109/TAC.2012.2195919}}

@article{Layden2016,
  title = {Universal scheme for indirect quantum control},
  author = {Layden, David and Mart\'{\i}n-Mart\'{\i}nez, Eduardo and Kempf, Achim},
  journal = {Phys. Rev. A},
  volume = {93},
  issue = {4},
  pages = {040301},
  numpages = {5},
  year = {2016},
  month = {Apr},
  publisher = {American Physical Society},
  doi = {10.1103/PhysRevA.93.040301},
  url = {https://link.aps.org/doi/10.1103/PhysRevA.93.040301}
}

@misc{taranto2025,
      title={Higher-Order Quantum Operations}, 
      author={Philip Taranto and Simon Milz and Mio Murao and Marco Túlio Quintino and Kavan Modi},
      year={2025},
      eprint={2503.09693},
      archivePrefix={arXiv},
      primaryClass={quant-ph},
      url={https://arxiv.org/abs/2503.09693}, 
}

@article{Chiribella2009,
  title = {Theoretical framework for quantum networks},
  author = {Chiribella, Giulio and D'Ariano, Giacomo Mauro and Perinotti, Paolo},
  journal = {Phys. Rev. A},
  volume = {80},
  issue = {2},
  pages = {022339},
  numpages = {20},
  year = {2009},
  month = {Aug},
  publisher = {American Physical Society},
  doi = {10.1103/PhysRevA.80.022339},
  url = {https://link.aps.org/doi/10.1103/PhysRevA.80.022339}
}

@ARTICLE{Oreshkov2012,
  title    = "Quantum correlations with no causal order",
  author   = "Oreshkov, Ognyan and Costa, Fabio and Brukner, {\v C}aslav",
  journal  = "Nature Communications",
  volume   =  3,
  number   =  1,
  pages    = "1092",
  month    =  oct,
  year     =  2012
}

@article{Costa_2016,
doi = {10.1088/1367-2630/18/6/063032},
url = {https://doi.org/10.1088/1367-2630/18/6/063032},
year = {2016},
month = {jun},
publisher = {IOP Publishing},
volume = {18},
number = {6},
pages = {063032},
author = {Costa, Fabio and Shrapnel, Sally},
title = {Quantum causal modelling},
journal = {New Journal of Physics}
}

@article{Pollock2018,
  title = {Operational Markov Condition for Quantum Processes},
  author = {Pollock, Felix A. and Rodr\'{\i}guez-Rosario, C\'esar and Frauenheim, Thomas and Paternostro, Mauro and Modi, Kavan},
  journal = {Phys. Rev. Lett.},
  volume = {120},
  issue = {4},
  pages = {040405},
  numpages = {6},
  year = {2018},
  month = {Jan},
  publisher = {American Physical Society},
  doi = {10.1103/PhysRevLett.120.040405},
  url = {https://link.aps.org/doi/10.1103/PhysRevLett.120.040405}
}

@article{Milz2021,
  title = {Quantum Stochastic Processes and Quantum non-Markovian Phenomena},
  author = {Milz, Simon and Modi, Kavan},
  journal = {PRX Quantum},
  volume = {2},
  issue = {3},
  pages = {030201},
  numpages = {81},
  year = {2021},
  month = {Jul},
  publisher = {American Physical Society},
  doi = {10.1103/PRXQuantum.2.030201},
  url = {https://link.aps.org/doi/10.1103/PRXQuantum.2.030201}
}

@article{Giarmatzi_2021,
   title={Witnessing quantum memory in non-Markovian processes},
   volume={5},
   ISSN={2521-327X},
   url={http://dx.doi.org/10.22331/q-2021-04-26-440},
   DOI={10.22331/q-2021-04-26-440},
   journal={Quantum},
   publisher={Verein zur Forderung des Open Access Publizierens in den Quantenwissenschaften},
   author={Giarmatzi, Christina and Costa, Fabio},
   year={2021},
   month=apr, pages={440} }

@ARTICLE{White2020,
  title    = "Demonstration of non-Markovian process characterisation and
              control on a quantum processor",
  author   = "White, G A L and Hill, C D and Pollock, F A and Hollenberg, L C L
              and Modi, K",
  journal  = "Nature Communications",
  volume   =  11,
  number   =  1,
  pages    = "6301",
  month    =  dec,
  year     =  2020
}

@article{Giarmatzi2025,
  doi = {10.22331/q-2025-12-18-1952},
  url = {https://doi.org/10.22331/q-2025-12-18-1952},
  title = {Multi-time quantum process tomography on a superconducting qubit},
  author = {Giarmatzi, Christina and Jones, Tyler and Gilchrist, Alexei and Pakkiam, Prasanna and Fedorov, Arkady and Costa, Fabio},
  journal = {{Quantum}},
  issn = {2521-327X},
  publisher = {{Verein zur F{\"{o}}rderung des Open Access Publizierens in den Quantenwissenschaften}},
  volume = {9},
  pages = {1952},
  month = dec,
  year = {2025}
}

@article{Figueroa-Romero2021,
  title = {Randomized Benchmarking for Non-Markovian Noise},
  author = {Figueroa-Romero, Pedro and Modi, Kavan and Harris, Robert J. and Stace, Thomas M. and Hsieh, Min-Hsiu},
  journal = {PRX Quantum},
  volume = {2},
  issue = {4},
  pages = {040351},
  numpages = {28},
  year = {2021},
  month = {Dec},
  publisher = {American Physical Society},
  doi = {10.1103/PRXQuantum.2.040351},
  url = {https://link.aps.org/doi/10.1103/PRXQuantum.2.040351}
}

@misc{srivastava2025,
      title={Blind-spots of Randomized Benchmarking Under Temporal Correlations}, 
      author={Varun Srivastava and Abhinash Kumar Roy and Soumik Mahanti and Jasleen Kaur and Salini Karuvade and Alexei Gilchrist},
      year={2025},
      eprint={2510.13051},
      archivePrefix={arXiv},
      primaryClass={quant-ph},
      url={https://arxiv.org/abs/2510.13051}, 
}

@misc{tanggara2024,
      title={Strategic Code: A Unified Spatio-Temporal Framework for Quantum Error-Correction}, 
      author={Andrew Tanggara and Mile Gu and Kishor Bharti},
      year={2024},
      eprint={2405.17567},
      archivePrefix={arXiv},
      primaryClass={quant-ph},
      url={https://arxiv.org/abs/2405.17567}, 
}

@article{White2025,
  title = {Unifying Non-Markovian Characterization with an Efficient and Self-Consistent Framework},
  author = {White, G. A. L. and Jurcevic, P. and Hill, C. D. and Modi, K.},
  journal = {Phys. Rev. X},
  volume = {15},
  issue = {2},
  pages = {021047},
  numpages = {43},
  year = {2025},
  month = {May},
  publisher = {American Physical Society},
  doi = {10.1103/PhysRevX.15.021047},
  url = {https://link.aps.org/doi/10.1103/PhysRevX.15.021047}
}

@article{Zambon2025,
  title = {Quantum Processes as Thermodynamic Resources: The Role of Non-Markovianity},
  author = {Zambon, Guilherme and Adesso, Gerardo},
  journal = {Phys. Rev. Lett.},
  volume = {134},
  issue = {20},
  pages = {200401},
  numpages = {9},
  year = {2025},
  month = {May},
  publisher = {American Physical Society},
  doi = {10.1103/PhysRevLett.134.200401},
  url = {https://link.aps.org/doi/10.1103/PhysRevLett.134.200401}
}

@inproceedings{Gutoski2007,
author = {Gutoski, Gus and Watrous, John},
title = {Toward a general theory of quantum games},
year = {2007},
isbn = {9781595936318},
publisher = {Association for Computing Machinery},
address = {New York, NY, USA},
url = {https://doi.org/10.1145/1250790.1250873},
doi = {10.1145/1250790.1250873},
booktitle = {Proceedings of the Thirty-Ninth Annual ACM Symposium on Theory of Computing},
pages = {565–574},
numpages = {10},
location = {San Diego, California, USA},
series = {STOC '07}
}

@article{Ziman2008,
  title = {Process positive-operator-valued measure: A mathematical framework for the description of process tomography experiments},
  author = {Ziman, M\'ario},
  journal = {Phys. Rev. A},
  volume = {77},
  issue = {6},
  pages = {062112},
  numpages = {4},
  year = {2008},
  month = {Jun},
  publisher = {American Physical Society},
  doi = {10.1103/PhysRevA.77.062112},
  url = {https://link.aps.org/doi/10.1103/PhysRevA.77.062112}
}

@article{Owari2015,
  title = {Probing an untouchable environment for its identification and control},
  author = {Owari, Masaki and Maruyama, Koji and Takui, Takeji and Kato, Go},
  journal = {Phys. Rev. A},
  volume = {91},
  issue = {1},
  pages = {012343},
  numpages = {20},
  year = {2015},
  month = {Jan},
  publisher = {American Physical Society},
  doi = {10.1103/PhysRevA.91.012343},
  url = {https://link.aps.org/doi/10.1103/PhysRevA.91.012343}
}

@article{CHOI1975,
title = {Completely positive linear maps on complex matrices},
journal = {Linear Algebra and its Applications},
volume = {10},
number = {3},
pages = {285-290},
year = {1975},
issn = {0024-3795},
doi = {https://doi.org/10.1016/0024-3795(75)90075-0},
url = {https://www.sciencedirect.com/science/article/pii/0024379575900750},
author = {Man-Duen Choi}
}

@article{JAMIOLKOWSKI1972,
title = {Linear transformations which preserve trace and positive semidefiniteness of operators},
journal = {Reports on Mathematical Physics},
volume = {3},
number = {4},
pages = {275-278},
year = {1972},
issn = {0034-4877},
doi = {https://doi.org/10.1016/0034-4877(72)90011-0},
url = {https://www.sciencedirect.com/science/article/pii/0034487772900110},
author = {A. Jamiołkowski}
}

@article{DOHERTY20131,
title = {The nitrogen-vacancy colour centre in diamond},
journal = {Physics Reports},
volume = {528},
number = {1},
pages = {1-45},
year = {2013},
note = {The nitrogen-vacancy colour centre in diamond},
issn = {0370-1573},
doi = {https://doi.org/10.1016/j.physrep.2013.02.001},
url = {https://www.sciencedirect.com/science/article/pii/S0370157313000562},
author = {Marcus W. Doherty and Neil B. Manson and Paul Delaney and Fedor Jelezko and Jörg Wrachtrup and Lloyd C.L. Hollenberg}
}

@book{Skrzypczyk_23,
author = {Skrzypczyk, Paul and Cavalcanti, Daniel},
title = {Semidefinite Programming in Quantum Information Science},
publisher = {IOP Publishing},
year = {2023},
series = {2053-2563},
isbn = {978-0-7503-3343-6},
url = {https://doi.org/10.1088/978-0-7503-3343-6},
doi = {10.1088/978-0-7503-3343-6}
}

@article{matsos2025,
  title = {Universal Quantum Gate Set for {{Gottesman}}--{{Kitaev}}--{{Preskill}} Logical Qubits},
  author = {Matsos, V. G. and Valahu, C. H. and Millican, M. J. and Navickas, T. and Kolesnikow, X. C. and Biercuk, M. J. and Tan, T. R.},
  year = 2025,
  month = oct,
  journal = {Nature Physics},
  volume = {21},
  number = {10},
  pages = {1664--1669},
  issn = {1745-2473, 1745-2481},
  doi = {10.1038/s41567-025-03002-8},
  urldate = {2026-06-23},
  langid = {english}
}

@misc{roy2026,
      title={Practical Tomography of Multi-Time Processes}, 
      author={Abhinash Kumar Roy and Varun Srivastava and Christina Giarmatzi and Alexei Gilchrist},
      year={2026},
      eprint={2604.01482},
      archivePrefix={arXiv},
      primaryClass={quant-ph},
      url={https://arxiv.org/abs/2604.01482}, 
}

@article{Chiribella_2016,
doi = {10.1088/1367-2630/18/9/093053},
url = {https://doi.org/10.1088/1367-2630/18/9/093053},
year = {2016},
month = {sep},
publisher = {IOP Publishing},
volume = {18},
number = {9},
pages = {093053},
author = {Chiribella, Giulio and Ebler, Daniel},
title = {Optimal quantum networks and one-shot entropies},
journal = {New Journal of Physics}
}

@article{White_2022,
  title = {Non-Markovian Quantum Process Tomography},
  author = {White, G.A.L. and Pollock, F.A. and Hollenberg, L.C.L. and Modi, K. and Hill, C.D.},
  journal = {PRX Quantum},
  volume = {3},
  issue = {2},
  pages = {020344},
  numpages = {30},
  year = {2022},
  month = {May},
  publisher = {American Physical Society},
  doi = {10.1103/PRXQuantum.3.020344},
  url = {https://link.aps.org/doi/10.1103/PRXQuantum.3.020344}
}

@article{Ortega_2024,
    author = {Ortega-Taberner, Carlos and O’Neill, Eoin and Butler, Eoin and Fux, Gerald E. and Eastham, P. R.},
    title = {Unifying methods for optimal control in non-Markovian quantum systems via process tensors},
    journal = {The Journal of Chemical Physics},
    volume = {161},
    number = {12},
    pages = {124119},
    year = {2024},
    month = {09},
    issn = {0021-9606},
    doi = {10.1063/5.0226031},
    url = {https://doi.org/10.1063/5.0226031},
}

@article{Degen2017,
  title = {Quantum sensing},
  author = {Degen, C. L. and Reinhard, F. and Cappellaro, P.},
  journal = {Rev. Mod. Phys.},
  volume = {89},
  issue = {3},
  pages = {035002},
  numpages = {39},
  year = {2017},
  month = {Jul},
  publisher = {American Physical Society},
  doi = {10.1103/RevModPhys.89.035002},
  url = {https://link.aps.org/doi/10.1103/RevModPhys.89.035002}
}

@article{Zhang_2019,
  title = {Improved Indirect Control of Nuclear Spins in Diamond N-$V$ Centers},
  author = {Zhang, Jingfu and Hegde, Swathi S. and Suter, Dieter},
  journal = {Phys. Rev. Appl.},
  volume = {12},
  issue = {6},
  pages = {064047},
  numpages = {11},
  year = {2019},
  month = {Dec},
  publisher = {American Physical Society},
  doi = {10.1103/PhysRevApplied.12.064047},
  url = {https://link.aps.org/doi/10.1103/PhysRevApplied.12.064047}
}

@article{Zhang_2011,
  title = {Coherent Control of Two Nuclear Spins Using the Anisotropic Hyperfine Interaction},
  author = {Zhang, Yingjie and Ryan, Colm A. and Laflamme, Raymond and Baugh, Jonathan},
  journal = {Phys. Rev. Lett.},
  volume = {107},
  issue = {17},
  pages = {170503},
  numpages = {5},
  year = {2011},
  month = {Oct},
  publisher = {American Physical Society},
  doi = {10.1103/PhysRevLett.107.170503},
  url = {https://link.aps.org/doi/10.1103/PhysRevLett.107.170503}
}

@article{Liu2013,
  author  = {Liu, G. Q. and Po, H. C. and Du, J. and Liu, R. B. and Pan, X. Y.},
  title   = {Noise-resilient quantum evolution steered by dynamical decoupling},
  journal = {Nature Communications},
  volume  = {4},
  pages   = {2254},
  year    = {2013},
  doi     = {10.1038/ncomms3254}
}

@article{London_2013,
  title = {Detecting and Polarizing Nuclear Spins with Double Resonance on a Single Electron Spin},
  author = {London, P. and Scheuer, J. and Cai, J.-M. and Schwarz, I. and Retzker, A. and Plenio, M. B. and Katagiri, M. and Teraji, T. and Koizumi, S. and Isoya, J. and Fischer, R. and McGuinness, L. P. and Naydenov, B. and Jelezko, F.},
  journal = {Phys. Rev. Lett.},
  volume = {111},
  issue = {6},
  pages = {067601},
  numpages = {5},
  year = {2013},
  month = {Aug},
  publisher = {American Physical Society},
  doi = {10.1103/PhysRevLett.111.067601},
  url = {https://link.aps.org/doi/10.1103/PhysRevLett.111.067601}
}

@article{Schwartz_2018,
author = {Ilai Schwartz  and Jochen Scheuer  and Benedikt Tratzmiller  and Samuel Müller  and Qiong Chen  and Ish Dhand  and Zhen-Yu Wang  and Christoph Müller  and Boris Naydenov  and Fedor Jelezko  and Martin B. Plenio },
title = {Robust optical polarization of nuclear spin baths using Hamiltonian engineering of nitrogen-vacancy center quantum dynamics},
journal = {Science Advances},
volume = {4},
number = {8},
pages = {eaat8978},
year = {2018},
doi = {10.1126/sciadv.aat8978},
URL = {https://www.science.org/doi/abs/10.1126/sciadv.aat8978},
eprint = {https://www.science.org/doi/pdf/10.1126/sciadv.aat8978}}

@article{Whaites_2023,
  title = {Hyperpolarization of nuclear spins: Polarization blockade},
  author = {Whaites, O. T. and Ioannou, C. I. and Pingault, B. J. and van de Stolpe, G. L. and Taminiau, T. H. and Monteiro, T. S.},
  journal = {Phys. Rev. Res.},
  volume = {5},
  issue = {4},
  pages = {043291},
  numpages = {14},
  year = {2023},
  month = {Dec},
  publisher = {American Physical Society},
  doi = {10.1103/PhysRevResearch.5.043291},
  url = {https://link.aps.org/doi/10.1103/PhysRevResearch.5.043291}
}

@article{Krantz_2019,
   title={A quantum engineer’s guide to superconducting qubits},
   volume={6},
   ISSN={1931-9401},
   url={http://dx.doi.org/10.1063/1.5089550},
   DOI={10.1063/1.5089550},
   number={2},
   journal={Applied Physics Reviews},
   publisher={AIP Publishing},
   author={Krantz, P. and Kjaergaard, M. and Yan, F. and Orlando, T. P. and Gustavsson, S. and Oliver, W. D.},
   year={2019},
   month=June }
